\documentclass[11pt]{article}
\pdfoutput=1
\usepackage{jheppub}
\usepackage{epstopdf}

\allowdisplaybreaks[4]

\usepackage{mathrsfs}
\usepackage{psfrag}
\usepackage{color}
\usepackage{hyperref}

\usepackage{slashed}
\usepackage{simplewick}
\usepackage{cancel}

\usepackage[table]{xcolor}

\usepackage{amsmath,bbm,array,amsfonts,graphicx,wrapfig,arydshln,lscape,float,multirow,longtable,rotating,makecell}
\usepackage{url}

\newcommand{\be}{\begin{equation}}
\newcommand{\ee}{\end{equation}}
\newcommand{\bea}{\begin{eqnarray}}
\newcommand{\eea}{\end{eqnarray}}
\newcommand{\Eq}[1]{Eq.~(\ref{#1})}

\newcommand{\Sec}[1]{Sec.~\ref{#1}}

\newcommand{\Fig}[1]{Figure~\ref{#1}}

\definecolor{forbidden}{rgb}{1,0.4,0.4}
\definecolor{trivial}{rgb}{0.4,0.4,1}

\definecolor{airforceblue}{rgb}{0.36, 0.54, 0.66}
\definecolor{bananayellow}{rgb}{1.0, 0.88, 0.21}
\definecolor{bittersweet}{rgb}{1.0, 0.44, 0.37}
\definecolor{blue(ncs)}{rgb}{0.0, 0.53, 0.74}
\definecolor{bole}{rgb}{0.47, 0.27, 0.23}
\definecolor{brass}{rgb}{0.71, 0.65, 0.26}
\definecolor{bronze}{rgb}{0.8, 0.5, 0.2}
\definecolor{brgreen}{rgb}{0.0, 0.26, 0.15}
\definecolor{burgundy}{rgb}{0.5, 0.0, 0.13}
\definecolor{cherry}{rgb}{1.0, 0.72, 0.77}
\definecolor{cocao}{rgb}{0.82, 0.41, 0.12}
\definecolor{citrine}{rgb}{0.99, 0.82, 0.07}

\title{\LARGE A Periodic Table of Effective Field Theories}

\author{Clifford Cheung,$^1$}
\author{Karol Kampf,$^2$}
\author{Jiri Novotny,$^2$}
\author{Chia-Hsien Shen,$^1$\hspace{3cm}}
\author{Jaroslav Trnka$^{3}$}

\affiliation{$^1$ Walter Burke Institute for Theoretical Physics,\\
California Institute of Technology, Pasadena, CA, USA}

\affiliation{$^2$ Institute of Particle and Nuclear Physics,\\
Faculty of Mathematics and Physics, Charles University, Prague, Czech Republic}

\affiliation{$^3$ Center for Quantum Mathematics and Physics (QMAP),\\ 
Department of Physics, University of California, Davis, CA, USA}

\emailAdd{clifford.cheung@caltech.edu, karol.kampf@mff.cuni.cz, jiri.novotny@mff.cuni.cz, chshen@caltech.edu, trnka@ucdavis.edu}

\abstract{We systematically explore the space of scalar effective field theories (EFTs) consistent with a Lorentz invariant and local S-matrix.  To do so we define an EFT classification based on four parameters characterizing 1) the number of derivatives per interaction, 2) the soft properties of amplitudes, 3) the leading valency of the interactions, and 4) the spacetime dimension.  Carving out the allowed space of EFTs, we prove that exceptional EFTs like the non-linear sigma model, Dirac-Born-Infeld theory, and the special Galileon lie precisely on the boundary of allowed theory space. Using on-shell momentum shifts and recursion relations, we prove that EFTs with arbitrarily soft behavior are forbidden and EFTs with leading valency much greater than the spacetime dimension cannot have enhanced soft behavior.  We then enumerate all single scalar EFTs in $d< 6$ and verify that they correspond to known theories in the literature.   
Our results suggest that the exceptional theories are the natural EFT analogs of gauge theory and gravity because they are one-parameter theories whose interactions are strictly dictated by properties of the S-matrix.
}

\preprint{
\begin{flushright}CALT-TH-2016-032\end{flushright}
}

\begin{document}

\maketitle 

\newpage

\section{Introduction}

The past couple decades have witnessed tremendous progress in our understanding of the S-matrix in gauge theory and gravity.  These developments have revealed hidden mathematical structures and symmetries that are completely invisible in the standard approach of Feynman diagrams. Moreover, they have led to alternative formulations of the S-matrix, for instance using recursion relations~\cite{Britto:2004ap,Britto:2005fq,ArkaniHamed:2008gz}, unitarity methods~\cite{Bern:1994cg,Bern:1994zx}, and more recently scattering equations~\cite{Cachazo:2013hca,Cachazo:2013iea,Cachazo:2014xea}, BCJ duality~\cite{Bern:2008qj,Bern:2010ue,Bern:2010yg}, hexagon bootstrap \cite{Dixon:2013eka,Caron-Huot:2016owq}, flux tube S-matrix \cite{Basso:2013vsa,Basso:2014hfa}, twistor methods~\cite{Witten:2003nn,Roiban:2004yf,Cachazo:2012kg,Cachazo:2012pz, ArkaniHamed:2009si,Mason:2013sva,Casali:2015vta,Geyer:2015bja,Geyer:2015jch}, Grassmannians~\cite{ArkaniHamed:2009dn}, on-shell diagrams and Amplituhedron~\cite{ ArkaniHamed:2012nw,Arkani-Hamed:2013jha,Arkani-Hamed:2013kca}.

While much of this work has centered on gauge theory and gravity, another important class of theories---effective field theories (EFTs)---have received substantially less attention, even though they play an important and ubiquitous role in many branches of physics.  At the very minimum, the EFT approach provides a general parameterization of dynamics in a particular regime of validity, usually taken to be low energies.  If the EFT has many free parameters then its predictive value is limited.  However, in many examples the interactions of the EFT are dictated by symmetries, {\it e.g}~as is the case for the Nambu-Goldstone bosons (NGBs) of spontaneous symmetry breaking.  At the level of scattering amplitudes, these rigid constraints are manifested by special infrared properties.  The archetype for this phenomenon is the Adler zero \cite{Adler:1964um},
\begin{equation}
\lim_{p\rightarrow0} A(p) = 0\,,
\end{equation}
which dictates the vanishing of amplitudes when the momentum of a NGB is taken to be soft. This imprint of symmetry on the S-matrix is reminiscent of gravity, which is also an EFT with a limited regime of validity.  

At the same time, the longstanding aim of the modern amplitudes program is to construct the S-matrix without the aid of a Lagrangian, thus relinquishing both the benefits and pitfalls of this standard approach.  But without a Lagrangian, it is far from obvious how to incorporate the symmetries of an EFT directly into the S-matrix.  However, recent progress in this direction~\cite{Cheung:2014dqa} has shown that the symmetries of many EFTs can be understood as the {\it consequence} of a ``generalized Adler zero'' characterizing a non-trivial vanishing of scattering amplitudes in the soft limit.  Here an amplitude is defined to have a ``non-trivial'' soft limit if it vanishes in the soft limit faster than one would naively expect given the number of derivatives per field.

By directly imposing a particular soft behavior at the level of the S-matrix, one can then {\it derive} EFTs and their symmetries from non-trivial soft behavior.  From this ``soft bootstrap'' one can rediscover a subclass of so-called ``exceptional'' EFTs~\cite{Cheung:2014dqa} whose leading interactions are uniquely fixed by a single coupling constant.  These exceptional theories include the non-linear sigma model (NLSM)~\cite{Cronin:1967jq,Weinberg:1966fm,Weinberg:1968de}, the Dirac-Born-Infeld (DBI) theory, and the so-called special Galileon~\cite{Cheung:2014dqa,Hinterbichler:2015pqa}.  

In~\cite{Cheung:2015ota}, it was shown the space of exceptional EFTs coincides precisely with the space of on-shell constructible theories via a new set of soft recursion relations.
These very same EFTs also appeared in a completely different context from the CHY scattering equations~\cite{Cachazo:2014xea}, which is a simple construction for building the S-matrices for certain theories of massless particles.  Altogether, these developments suggest that the exceptional theories are the EFT analogs of gauge theory and gravity.  In particular, they are all simple one-parameter theories whose interactions are fully fixed by  simple properties of the S-matrix.

\begin{figure}[t]
	\begin{center}
		\includegraphics[height=0.55\textwidth]{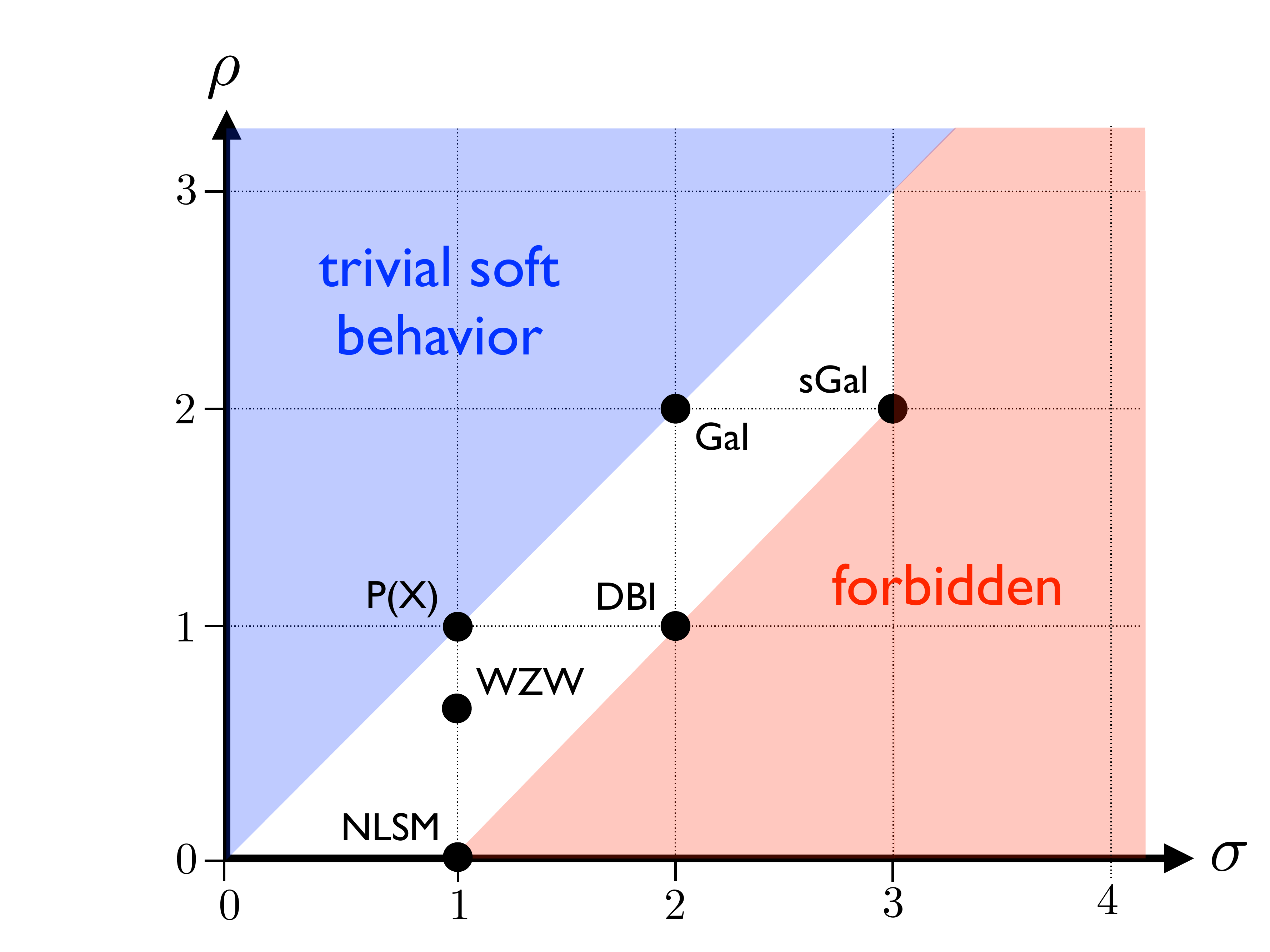}
	\end{center}
	\vspace{-5mm}
	\caption{
		Plot summarizing the allowed parameter space of EFTs.
		The blue region denotes EFTs whose soft behavior is trivial due to the number of derivatives per interaction.  The red region is forbidden by consistency of the S-matrix, as discussed in \Sec{sec:theory_space}.  The white region denotes EFTs with non-trivial soft behavior, with solid black circles representing known standalone theories. The $d$-dimensional WZW term theory corresponds to $(\rho,\sigma)=(\frac{d-2}{d-1},1)$.  The exceptional EFTs all lie on the boundary of allowed theory space and $(\rho,\sigma)=(3,3)$ is forbidden.}
	\label{fig:EFT_map}
\end{figure}

In this paper, we systematically carve out the theory space of all possible Lorentz invariant and local scalar EFTs by imposing physical consistency conditions on their on-shell scattering amplitudes.  Our classification hinges on a set of physical parameters $(\rho,\sigma, v,d)$ which label a given hypothetical EFT.  Here $\rho$ characterizes the number of derivatives per interaction, with a corresponding Lagrangian of the schematic form
\begin{equation}
{\cal L} = \partial^2 \phi^2 F(\partial^\rho \phi)\,,
\label{eq:def_rho}
\end{equation}
for some function $F$.
This power counting structure is required for destructive interference between tree diagrams of different topologies \cite{Cheung:2014dqa}. Meanwhile, the parameter $\sigma$ is the soft degree characterizing the power at which amplitudes vanish in the soft limit, 
\begin{equation}
\lim_{p\rightarrow0} A(p) = {\cal O}(p^\sigma)\,.
\label{eq:def_sigma}
\end{equation}
Obviously, for sufficiently large $\rho$, a large of value $\sigma$ is trivial because a theory with many derivatives per field will automatically have a higher degree soft limit.  As shown in \cite{Cheung:2014dqa} the soft limit becomes non-trivial when
\bea
\sigma \geq \rho& \quad \textrm{for} \quad \rho >1\,, \nonumber \\
\sigma > \rho& \quad \textrm{for} \quad \rho \leq 1 \,.
\label{eq:triviality}
\eea
The other parameters in our classification are $v$, the valency of the leading interaction, and $d$,  the space-time dimension.

Taking a bottom up approach, we {\it assume} a set of values for $(\rho,\sigma, v,d)$ to bootstrap scattering amplitudes which we then analyze for self-consistency.  Remarkably, by fixing these parameters---without the aid of a specific Lagrangian or set of symmetries---it is possible to rule out whole swaths of EFT space using only properties of the S-matrix.  Since our analysis sidesteps top down considerations coming from symmetries and Lagrangians, we obtain a robust system for classifying and excluding EFTs.  This approach yields an overarching organizing principle for EFTs, depicted pictorially in \Fig{fig:EFT_map} as a sort of ``periodic table'' for these structures.  See Appendix~\ref{app:sumoftheories} for a brief summary of the EFTs discussed in this paper.
Our main results are as follows:
\begin{itemize}

\item The soft degree of all EFTs is bounded by the number of derivatives per interaction, so in particular, $\sigma \leq \rho+ 1$.  The exceptional EFTs---the NLSM, DBI, and the special Galileon---all saturate this bound.

\item The soft degree of every non-trivial EFT is strictly bounded by $\sigma \leq 3$, so arbitrarily enhanced soft limits are forbidden.

\item Non-trivial soft limits require the valency of the leading interaction be bounded by the spacetime dimension, so $v \leq d+1$.  For $4 < v \leq d+1$, this is saturated by the Galileon~\cite{Horndeski:1974wa,Nicolis:2008in} and the Wess-Zumino-Witten (WZW) term for the NLSM~\cite{WZWterm1,WZWterm2}.

\item The above constraints permit a theory space of single scalar EFTs and multiple scalar EFTs with flavor-ordering in general $d$ populated by known theories: NLSM, DBI, the Galileon, and WZW.  In principle this allows for new theories at the these same values of $(\rho,\sigma,d,v)$ but we exclude this possibility in $d=3,4,5$ by direct enumeration.  
\end{itemize}

\noindent The core results of this paper focus on the soft behavior of EFTs of a single scalar, or multiple scalars where there is a notion of flavor-ordering.  However, we also briefly discuss the space of general EFTs with multiple scalars, as well as alternative kinematical regimes like the double soft or collinear limits.  

The paper is organized as follows.  In \Sec{sec:scheme}, we define the parameters of the EFT theory space and outline our strategy for classification. 
We then derive soft theorems from general symmetry considerations in \Sec{sec:symm}. 
The tools for classification---soft momentum shifts and recursion relations---are summarized in \Sec{sec:recursions}, and then applied to carve out the space of allowed EFTs in \Sec{sec:theory_space}. In the permitted region, we search and enumerate EFTs numerically in \Sec{sec:enumeration}.
Other kinematics limits and more general classes of theories are considered in \Sec{sec:more}. Finally we conclude in \Sec{sec:conclusion}.

\section{Classification scheme}
\label{sec:scheme}

As described in the introduction, scalar EFTs are naturally classified in terms of the set of parameters $(\rho,\sigma,v,d)$. 
Here we review the definitions and motivations for these parameters, first in terms of the Lagrangian and then in terms of the S-matrix.

\subsection{Lagrangians}

The power counting parameter $\rho$ is a measure of the number of powers of momentum associated with each interaction. 
As shown in~\cite{Cheung:2014dqa}, destructive interference among diagrams, {\it i.e.} cancellations, imposes a strict power counting condition relating the interactions of the EFT.
In particular, suppose that the Lagrangian has a schematic form
\begin{equation}
{\cal L} = \sum\limits_{m=0}^\infty \sum\limits_{n=v}^\infty \lambda_{m,n} \; \partial^{m} \phi^{n}\,, 
\label{eq:L}
\end{equation}
where $\lambda_{m,n}$ are coupling constants.
Cancellations can occur between couplings of fixed
\begin{equation}
\rho =\frac{m-2}{n-2}\,, \label{eq:rho_def}
\end{equation}
where $\rho$ is a fixed non-negative rational number.   Here \Eq{eq:L} is schematic, since we have suppressed Lorentz and internal indices so at a given order in $m,n$ there are actually many coupling constants $\lambda_{m,n}$.  This restriction still leaves a huge parameter space of viable EFTs.

In principle, one can combine interactions of different values of $\rho$ into the same theory.  However, cancellations among the interactions with either the smallest or the highest value of $\rho$ are closed, so it is natural to focus first on fixed $\rho$ theories.

In \Eq{eq:L}, $v$ denotes the valency of the leading interaction.  Naively, the minimal possible valency is $v=3$.  However, the leading cubic vertex in a derivatively coupled theory of massless scalars can always be eliminated by equations of motion.  This is obvious because the only possible non-zero 3pt amplitude of scalars is a constant, corresponding to a cubic scalar potential interaction.  On the other hand, the on-shell 3pt amplitude for derivatively coupled scalars will vanish because there is no non-zero kinematic invariant built from three on-shell momenta.   So without loss of generality we can take $v=4$ as the minimum valency.

For concreteness, let us briefly enumerate a few simple examples of Lagrangians with fixed $\rho$. Consider first the very simplest case, $\rho=0$, for a theory of a single scalar with only even interactions,
\begin{equation}
{\cal L}_{\rho=0} = \lambda_{2,4} (\partial^2\phi^4) + \lambda_{2,6} (\partial^2\phi^6) + \lambda_{2,8} (\partial^2\phi^8) + \dots
\end{equation}
Since each term only has two derivatives, the Lorentz structure of these terms is simple: 
\begin{equation}
\phi^{n-2}(\partial^\mu\phi\,\partial_\mu\phi)\,.
\end{equation}
It is straightforward to see that all on-shell tree-level scattering amplitudes in this theory are zero, corresponding to the fact that all the interactions are related by a field redefinition to the action for a free scalar.  For a multiplet of scalars, this is no longer true, and the theory can have non-trivial scattering amplitudes.  

For $\rho=1$ the Lagrangian for a scalar with even interactions is
\begin{equation}
{\cal L}_{\rho=1} = \lambda_{4,4} (\partial^4\phi^4) + \lambda_{6,6} (\partial^6\phi^6) + \lambda_{8,8} (\partial^8\phi^8) + \dots
\label{eq:rho1}
\end{equation}
In this case, even for a single scalar field there are many possible ways to contract Lorentz indices.  For example, the first term above could represent any of three different interactions, 
\begin{equation}
\lambda_{4,4}^{(1)} (\partial^\mu\phi)(\partial_\mu\phi) (\partial^\nu\phi)(\partial_\nu\phi) + \lambda_{4,4}^{(2)}\phi^2(\partial^{\mu}\partial^{\nu}\phi)(\partial_{\mu}\partial_{\nu}\phi) + \lambda_{4,4}^{(3)}\phi(\partial^{\mu}\partial^\nu\phi)(\partial_{\mu}\phi)(\partial_\nu \phi) \,.
\end{equation}
In fact, we can eliminate two of these terms via integration-by-parts identities and equations of motion. These relations are harder to track down for more complicated Lagrangians, but for our analysis we will thankfully not need to determine all of these identities.

Finally, let us stress that $\rho$ need not be an integer, but is more generally an arbitrary rational number. As we will later see, a case of particular interest is $\rho=2/3$, for which
\begin{equation}
{\cal L}_{\rho=\frac{2}{3}} = \lambda_{4,5} (\partial^4\phi^5) + \lambda_{6,8} (\partial^6\phi^8) + \lambda_{8,11} (\partial^8\phi^{11}) + \dots
\end{equation}
A priori, quite extreme values of $\rho$ are possible.  For example, for $\rho=13/11$ we have
\begin{equation}
{\cal L}_{\rho=\frac{13}{11}} = \lambda_{28,24} (\partial^{28}\phi^{24}) + \lambda_{54,46} (\partial^{54}\phi^{46})  + \dots
\end{equation}
For such peculiar values of $\rho$, the leading valency $v$ of the theory can be very high.  Naively, this signals a serious obstruction to any program for explicit construction of all possible EFTs.  In particular, any exhaustive search for EFTs at a fixed valency will always miss possible EFTs at higher valency.  After all, the space of rational numbers $\rho$ is dense.  Remarkably, we will later on find general arguments bounding the allowed maximum valency of a consistent EFT, making an enumerative procedure feasible.  

Although only theories with fixed $\rho$ are considered in this paper, we briefly comment on the scenario with multiple $\rho$ interactions. This generally arises from loop induced interactions. For instance, the 1-loop correction of Eq.~\eqref{eq:rho1} yields
\begin{equation}
{\cal L}' = \lambda_{8,4} (\partial^8\phi^4) + \lambda_{10,6} (\partial^{10}\phi^6) + \lambda_{12,8} (\partial^{12}\phi^8) + \dots
\end{equation}
The single insertion of the above operators corresponds to $\rho=3,2,5/3$ for four, six, and eight points respectively. Given fixed loop order counting, we find the value of $\rho$ decreases for higher point interactions. Suppose the associated amplitudes have soft limit $\sigma=2$ (which we expect for the loop-correction of DBI theory).
The amplitudes will have trivial soft limits at four point but become non-trivial starting at six point. We leave the study of multiple $\rho$ theories to future work.

\subsection{Scattering Amplitudes}

Starting from a general Lagrangian of fixed power counting parameter $\rho$ one can calculate the $n$pt tree-level scattering amplitude using the corresponding Feynman rules. The resulting answer is a function of kinematical invariants together with the coupling constants $\lambda_{m,n}$.   In turn, the $\lambda_{m,n}$ can be constrained by demanding that the amplitude conform to the enhanced soft limit of \Eq{eq:def_sigma}.

In principle, the soft degree $\sigma$ can be any integer.  However, $\sigma<0$ corresponds to singular behavior in the soft limit, which is only possible if there are cubic interactions in the theory.  As we argued previously, though, all such cubic interactions can be eliminated by equations of motion in a theory of derivatively coupled scalars.  In contrast, such cubic interactions are physical in YM and gravity, where $\sigma = -1$.  In any case, for scalar EFTs we have that $\sigma \ge 0$.

As the number of derivatives per field increases, so too will the soft degree.  However, something interesting occurs when the soft degree exceeds the number of derivatives per field,
\begin{equation}
	\sigma > \frac{m}{n}\,,
\end{equation}
which is only possible if there is cancellation among diagrams. We define this to be an enhanced soft limit (see \cite{Cheung:2014dqa}). Rewriting this inequality in terms of $\rho$, we obtain
\begin{equation}
	(\sigma-1) > (\rho -1) \times \left(1-\frac{2}{n}\right)\,.
\end{equation}
For a theory with enhanced soft behavior, this inequality should be true of all amplitudes.  Thus we can take the large $n$ limit, in which case the inequality approaches the inequalities in \Eq{eq:triviality}.
This range defines a swath of EFT space that has enhanced soft behavior, which will be of our primary interest.

\subsection{Ansatze}

Fixing the power counting parameter $\rho$, the soft degree $\sigma$, the valency of the leading interaction $v$, and the spacetime dimension $d$, we can now place stringent constraints on the space of scalar EFTs.  One way to  compute the associated scattering amplitudes would be natural to enumerate all possible Lagrangian terms and calculate using Feynman diagrams. While this approach is straightforward, it is plagued with redundancies since integration-by-parts identities and field redefinitions induce an infinite set of Lagrangians corresponding to identical physics. Indeed, even a systematic enumeration of higher dimension operators in EFTs is a non-trivial task that remains an active area of research~\cite{Henning:2015alf}.

Here we bypass this complication by directly constructing the scattering amplitudes using ansatze.  For a theory of scalars, the tree-level scattering amplitude $A_n$ is a rational function of kinematic invariants $s_{ij} = (p_i+p_j)^2$,
where $A_n$ has poles only when $s_{i_1i_2\dots i_k} = (p_{i_1}+p_{i_2}+\dots+p_{i_k})^2=0$.
Note the absence of two particle poles, $s_{ij}=0$, since the 3pt amplitude vanishes in a theory of derivatively coupled scalars.  Schematically, the scattering amplitude ansatz is
\begin{equation}
A_{n,m}(s_{ij}) = 
\sum_{\rm topology} \frac{N(s_{ij})}{D(s_{ij})} + A_{\rm contact}(s_{ij})\,,
\label{eq:amp_ansatz}
\end{equation}
where $m=\rho(n-2)+2$ is the dimension of the amplitude, and counts the net power of momenta in the amplitude.  Here the summation runs over all topologies involving internal exchanged scalars, allowing for all possible interactions consistent with $\rho$.  These terms enter with propagator denominators collected into the function $D$, and the remaining numerator function is $N$.  The second term $A_{\rm contact}$ corresponds to contributions that do not have propagator denominators, and is thus a local function of the kinematic invariants.

The amplitudes ansatz should satisfy several consistency conditions.   First, it must factorize properly on poles, so
\begin{equation}
\lim_{P^2\rightarrow0} A_{n,m} = \sum \frac{A_L A_R}{P^2}\,,
\label{eq:fact}
\end{equation}
where $P=(p_{i_1}+p_{i_2}+\dots + p_{i_k})$ and the sum runs over internal states. 
Second, the amplitudes ansatz should respect all the permutation symmetries of a given diagram. For example, in a theory of a single scalar, all vertices should be permutation invariant under the exchange of external legs and all diagrams of the same topology should be related by permutations.  

An ansatz consistent with the above conditions is a genuine scattering amplitude corresponding to the conjugacy class of physically equivalent Lagrangians that are identical up to off-shell redundancies like field redefinitions and integration-by-parts identities.  The immense advantage of these amplitudes ansatze is that these objects are free from such off-shell ambiguities and thus uniquely label distinct theories.

To be concrete, let us spell out this ansatz construction explicitly for the 4pt and 6pt amplitudes for a $\rho=1$ theory.  The unique 4pt amplitude for such a theory is
\begin{equation}
	A_4 = \lambda_{4,1} (s_{12}^2+s_{13}^2+s_{23}^2)\,.
\end{equation}
Since there is only one possible invariant, the corresponding Lagrangian must only describe one physical interaction parameterized by $\lambda_{4,1}$  

The 6pt amplitudes ansatz has a contact term and ten factorization terms
\begin{equation}
A_6=  \left(\frac{\lambda_{4,1}^2(s_{12}^2+s_{13}^2+s_{23}^2)(s_{45}^2+s_{46}^2+s_{56}^2)}{s_{123}} +{\rm permutations}\right)+A_{6,{\rm contact}}\,,
\end{equation}
where the permutations run through the other nine factorization channels.
The factorization term is written so as to factorizes properly into 4pt amplitudes while the contact term is
\begin{align}
A_{6,{\rm contact}} &= \alpha_1  s_{12}^3 + \alpha_2  s_{12}^2s_{13} + \alpha_3  s_{12}^2s_{34} + \alpha_4   s_{12}s_{13}s_{23}+ \alpha_5 s_{12}s_{13}s_{14} + \alpha_6  s_{12}s_{23}s_{34} \nonumber\\ 
&\hspace{1cm}+ \alpha_7   s_{12}s_{23}s_{45} + \alpha_8  s_{12}s_{34}s_{56} + \mbox{symmetrization in (123456)}\,.
\end{align}
Not all these terms are independent, but kinematical identities eliminate all but two terms which can be chosen to be the terms proportional to $\alpha_1$, $\alpha_2$, $\alpha_4$, $\alpha_5$.  

In general, it is difficult to enumerate all of these kinematical identities analytically in order to reduce the ansatz to an independent basis of terms. Such a task is essentially equivalent to identifying an independent set of Lagrangian operators. However, by working with the ansatz directly, we can evaluate the ansatz numerically in order to remove the elements that generate numerically identical amplitudes.

Lastly, we note that in analogy with color-ordering in YM theory, it is sometimes possible to cleanly disaggregate the Lie algebraic and kinematic elements of the amplitude in an EFT of multiple scalars.  For example, in the NLSM, a scattering amplitude $A_n$ can be written as a sum over flavor-ordered amplitudes~\cite{Kampf:2012fn,Kampf:2013vha}
\begin{equation}
A_{n} = \sum_{S/Z_n} {\rm Tr}(T^{\sigma_{a_1}}T^{\sigma_{a_2}}\dots T^{\sigma_{a_n}})\,A_n^{(s)}(\sigma_{a_1},\sigma_{a_2},\dots \sigma_{a_n})\,.
\label{strip}
\end{equation}
After stripping off the Lie algebra structure, the flavor-ordered amplitudes are cyclically invariant with poles only in adjacent factorization channels like $s_{123}=0$ and $s_{2345}=0$.  For these flavor-ordered amplitudes, the procedure for contracting ansatze is the same as before, only subject to the extra conditions of adjacent factorization and cyclic symmetry.

A priori, flavor ordering is not always possible in a general EFT of multiple scalars.  In certain cases the flavor decomposition will involve multitrace terms, even in the tree-level scattering amplitude.  While the bulk of this paper is focused on the amplitudes for scalar field or the flavor-ordered amplitudes for multiple scalars, in \Sec{sec:multiplescalars} we also discuss some results for genuine multiple scalar field theories where the flavor-ordering is not assumed. 

\section{From Symmetries to Soft Limits}
\label{sec:symm}

In this section we revisit the traditional field theory approach whereby the
soft limit is derived a byproduct of symmetry. From this
perspective the vanishing of scattering amplitudes---\textit{e.g.}~the
so-called Adler zero of NGBs---arises from spontaneous symmetry breaking in
the EFT. Here the key observation is that the scattering amplitude of a soft
NGB is closely related to the matrix element of the corresponding Noether
current $J^{\mu }$, in particular with a certain regular remainder function $R^{\mu }(p)$ obtained when the
one-particle pole of the soft NGB is subtracted (cf.~\Eq{remnant_formula}) below\footnote{%
For further details see \textit{e.g.}~the textbook~\cite{Weinberg:1996kr}
and references therein.}). Therefore, the soft behavior of amplitudes is dictated by the properties of the Noether currents of
spontaneously broken symmetries.

The EFTs we consider here are derivatively coupled. Most of them are invariant
with respect to the simple shift symmetry, 
\begin{equation}
\phi (x)\rightarrow \phi (x)+a  \label{shift0}\,,
\end{equation}%
which is spontaneously broken, yielding a corresponding NGB field $\phi$.  Provided we have additional information on the
Noether current of the shift symmetry at our disposal, we can further
deduce  soft limit properties of the scattering amplitudes beyond the leading Adler zero. This additional information is
obtained from the enhanced symmetries of the theory.

While the technical steps of the subsequent analysis are somewhat
complicated, our final conclusion is quite simple. In
order to obtain an enhanced $\mathcal{O}\left( p^{n+1}\right) $ soft
behavior of the amplitudes, it is sufficient that there is an additional
non-linear (\textit{i.e.} spontaneously broken) symmetry of the action of
the form 
\begin{equation}
\delta \phi \left( x\right) =\theta _{\alpha _{1}\ldots \alpha _{n}}\left[
x^{\alpha _{1}}\ldots x^{\alpha _{n}}+\Delta ^{\alpha _{1}\ldots \alpha
_{n}}\left( x\right) \right]\,,  \label{sym}
\end{equation}%
where $\Delta ^{\alpha _{1}\ldots \alpha _{n}}\left( x\right) $ is linear
combination of local composite operators comprised of $\phi$ with coefficients that have polynomial
dependence on $x$.\ More precisely, under some regularity assumptions (%
\textit{e.g.}. absence of cubic vertices), and (almost) irrespectively on
the explicit form of $\Delta ^{\alpha _{1}\ldots \alpha _{n}}\left( x\right) 
$, the very presence of the symmetry in \Eq{sym} is sufficient condition for
the $\mathcal{O}\left( p^{n+1}\right) $ behavior of the resulting scattering
amplitudes corresponding to $\sigma =n+1$. Let us note that this result
depends only on the $c-$number part of the general symmetry transformation
Eq.~(\ref{sym}). Therefore, theories invariant with respect to the
transformation in \Eq{sym} with the same polynomial $\alpha ( x)
=\theta _{\alpha _{1}\ldots \alpha _{n}}x^{\alpha _{1}}\ldots x^{\alpha _{n}}
$ form a universality class of the same soft behavior.

We relegate the details of our proof to Appendix \ref{app:proofsoft}, but here simply sketch the main steps of the argument.  A crucial ingredient of
the proof is an observation that the Noether currents of the shift symmetry
and of the enhanced symmetry in \Eq{sym} are in fact closely related (for
more details see \cite{Brauner:2014aha}). For single scalar EFTs this
can be easily understood intuitively: there is only one NGB (which
corresponds to the shift symmetry) but more than one non-linear (i.e.
spontaneously broken) symmetry; thus the Noether currents cannot be
independent. At the classical level there is another more precise
argument.  When we promote the global symmetries to local ones (i.e. when the
parameters $a$ and $\theta _{\alpha _{1}\ldots \alpha _{n}}$ become
space-time dependent), the localized symmetry in \Eq{sym} can be treated as
a localized shift symmetry \Eq{shift0} with very special parameter%
\begin{equation}
a\rightarrow \widehat{a}\left( x\right) =\theta _{\alpha _{1}\ldots \alpha
_{n}}\left( x\right) \left[ x^{\alpha _{1}}\ldots x^{\alpha _{n}}+\Delta
^{\alpha _{1}\ldots \alpha _{n}}\left( x\right) \right] .
\end{equation}%
The above relations between
currents express the Noether currents of the symmetry \Eq{sym}
in terms of the shift symmetry current $J^{\mu }$, and more importantly put a constraint
on the possible form of $J^{\mu }$ itself. At the quantum level%
\footnote{%
Such a relation holds automatically at tree-level and we assume here that is not spoiled by the quantum corrections.} this constraint reads 
\begin{equation}
\left\langle \alpha ,\mathrm{out}|J^{\mu }\left( x\right) |\beta ,\mathrm{in}%
\right\rangle \partial _{\mu }x^{\alpha _{1}}\ldots x^{\alpha _{n}}=\partial
_{\mu }\left\langle \alpha ,\mathrm{out}|\Gamma ^{\mu \alpha _{1}\ldots
\alpha _{n}}\left( x\right) |\beta ,\mathrm{in}\right\rangle \,,
\label{current_relation_0}
\end{equation}
where $\Gamma ^{\mu \alpha _{1}\ldots \alpha _{n}}\left( x\right) =\gamma
_{A}^{\mu \alpha _{1}\ldots \alpha _{n}}\left( x\right) O^{A}\left( x\right) 
$ is some linear combination of local composite operators $O^{A}\left(
x\right) $ with coefficients $\gamma _{A}^{\mu \alpha _{1}\ldots \alpha
_{n}}\left( x\right) $ with polynomial dependence on $x$. The explicit form of $\Gamma ^{\mu \alpha _{1}\ldots \alpha _{n}}\left(
x\right) $ which depends on $\Delta ^{\alpha _{1}\ldots \alpha _{n}}\left(
x\right) $ is irrelevant for the proof of the soft theorem.

Subtracting the one-particle pole in $p$ (where $p=P_{\beta }-P_{\alpha }$ is a difference of momenta in the $in$ and $out$
state) on both sides of the relation in \Eq{current_relation_0}, we obtain a relation between the regular remainder function $R^{\mu }\left( p\right) $
of the matrix element of the shift current and the regular remainders $%
R^{A}\left( p\right) $ of the local operators $O^{A}\left( x\right) $. Such
a relation reads
\begin{equation}
e^{-ip\cdot x}\partial _{\mu }x^{\alpha _{1}}\ldots x^{\alpha _{n}}R^{\mu
}(p)=\partial _{\mu }\left[ \gamma _{A}^{\mu \alpha _{1}\ldots \alpha
_{n}}\left( x\right) e^{-ip\cdot x}\right] R^{A}(p)\,.
\end{equation}
Assuming regularity\footnote{%
Regularity of $R^{\mu }$ for $p\rightarrow 0$ is guaranteed in the absence
of the cubic vertices.} 
of the remainders for $p\rightarrow 0$, we can
integrate over $\mathrm{d}^{d}x$ to obtain
\begin{equation}
p_{\mu }R^{\mu }(p)\partial ^{\alpha _{1}}\ldots \partial ^{\alpha
_{n}}\delta ^{(4)}(p)=0\,.  \label{soft_theorem_formula}
\end{equation}%
The latter formula, together with 
\begin{equation}
\left\langle \alpha +\phi (\mathbf{p}),\mathrm{out}|\beta ,\mathrm{in}%
\right\rangle =\frac{1}{F}p_{\mu }R^{\mu }(p)\,,  \label{remnant_formula}
\end{equation}
which relates the remainder function to the NGB amplitude is at the core of the
soft theorems for theories with the enhanced symmetry in \Eq{sym}.

As an example let us consider theories which belong to the universality
class of theories invariant with respect to Eq. (\ref{sym}) for which 
\begin{equation}
\alpha \left( x\right) \propto \theta \cdot x\,.
\end{equation}
Prominent members of this class are the general Galileon and DBI. While the former is invariant
with respect to the linear shift 
\begin{equation}
\delta _{\theta }\phi \left( x\right) =\theta \cdot x\,,
\end{equation}
the latter has a nonlinearly realized $\left( d+1\right) $-dimensional
Lorentz symmetry 
\begin{equation}
\delta _{\theta }\phi \left( x\right) =\theta \cdot x-F^{-d}\theta \cdot
\phi (x)\partial \phi \left( x\right) .
\end{equation}
Inserting the above $\alpha \left( x\right)$ into Eq.~(\ref{soft_theorem_formula}) we get
\begin{eqnarray}
0 &=&p_{\mu }R^{\mu }(p)\partial ^{\beta }\delta ^{(d)}(p)  \nonumber \\
&=&-\left[ \partial ^{\beta }\delta ^{(d)}(p)\right] \left[
\lim_{p\rightarrow 0}p_{\mu }R^{\mu }(p)\right] -\delta ^{(d)}(p)\left[
\lim_{p\rightarrow 0}\partial ^{\alpha }\left( p_{\mu }R^{\mu }(p)\right) %
\right] \,.
\label{linear_shift_class}
\end{eqnarray}
We recover thus not only the Adler zero condition 
\begin{equation}
\lim_{p\rightarrow 0}p_{\mu }R^{\mu }(p)=0\,,
\end{equation}
but also an enhanced $\mathcal{O}\left( p^{2}\right) $ soft behavior
corresponding to
\begin{equation}
\lim_{p\rightarrow 0}\partial ^{\alpha }\left( p_{\mu }R^{\mu }(p)\right)
=F\lim_{p\rightarrow 0}\partial ^{\alpha }\left\langle \alpha +\phi (\mathbf{%
p}),\mathrm{out}|\beta ,\mathrm{in}\right\rangle =0\,,
\end{equation}
implying an Adler zero of the second degree. Further applications and
generalizations can be found in Appendix~\ref{app:proofsoft}.

\section{On-shell Reconstruction}
\label{sec:recursions}

As demonstrated in \cite{Cheung:2014dqa}, enhanced soft behavior can be sufficiently constraining so as to fully dictate all tree amplitudes up to a single coupling constant.  So for these exceptional EFTs,  soft limits and factorization are enough information to fully determine the S-matrix.  Since these EFTs are so special, they naturally reside near the boundary of the allowed regions of EFT space, which we verify explicitly in Sec.~\ref{sec:theory_space}.

In the present section, we introduce the notion of on-shell constructibility, which is critical for bootstrapping the S-matrix of a given EFT.  The concept of on-shell constructibility arose originally in YM theory and gravity, where tree-level amplitudes are fully fixed by two conditions: gauge invariance and factorization.  The factorization condition, shown in \Eq{eq:fact}, can then be imposed sequentially until all higher point amplitudes are reduced in terms of a set of input 3pt amplitudes.  Said another way, the physical  $n$-pt amplitude is the unique gauge invariant function which satisfies \Eq{eq:fact} in all channels.  

Conveniently, the dual conditions of gauge invariance and factorization can be imposed automatically in YM and gravity using the celebrated BCFW recursion relations \cite{Britto:2004ap,Britto:2005fq}.  These work by applying a complex shift of the momenta,
\begin{equation}
p_i \rightarrow p_i + zq\,,\qquad p_j \rightarrow p_j - zq\,,
\end{equation}
where $q^2=(p_i\cdot q)=(p_j\cdot q)=0$ and the momentum conservation is preserved. Applying Cauchy's formula to the shifted amplitude $A_n(z)$, we can then reconstruct the original $A_n$ using the products of shifted lower point amplitudes,
\begin{equation}
\int \frac{dz\,A_n(z)}{z} = 0\quad\rightarrow\quad A_n =  \sum_{k} \frac{A_L(z_k)A_R(z_k)}{P^2}\,,
\label{Cauchy2}
\end{equation}
where the sum is over all factorization channels for which $P^2(z_k)=0$. Later on, the BCFW recursion relations were generalized to apply to a much broader class of theories~\cite{Cheung:2008dn,Cohen:2010mi,Cheung:2015cba}.

An important requirement of Eq.~\eqref{Cauchy2} is that the shifted amplitude falls off at infinity, $A_n(z)\sim \frac{1}{z}$ for $z\rightarrow\infty$. If this is not true, then the recursion includes boundary terms which are difficult to calculate, though some progress has been recently made on that front \cite{Feng:2014pia,Jin:2014qya,Jin:2015pua}. For EFTs, amplitudes typically grow at large $z$ as $A_n(z)\sim z^p$ where $p>0$, so none of the standard recursion relations can be used. 

This obstruction to recursion in EFTs is obvious from a physical perspective: typically there is an infinite tower of interactions in EFTs which produces contact terms in amplitudes.
These contact terms cannot be constrained by factorization. 
So we need additional information to fix these unconstrained contact terms.
In YM and gravity, gauge invariance dictates the appearance of contact terms and makes reconstruction feasible.  In principle, it may be possible that these contact interactions can be fixed by leading and subleading soft theorems, and in particular recent work on conformal field structures for amplitudes suggest this may occur \cite{Cheung:2016iub}.

In scalar EFT, there is no gauge invariance to speak of, so it is natural to consider soft structure to relate cancellations between contact and pole terms.
In particular, we call the amplitude $A_n$ {\it soft limit constructible} if it is the unique function satisfying two conditions:
\begin{enumerate}
	\item It has local poles and factorizes correctly on them according to Eq.~\eqref{eq:fact}.
	\item It has required soft limit behavior $A_n={\cal O}(p^\sigma)$. 
\end{enumerate}
Soft limit constructibility imposes non-trivial conditions on our classification parameters $(\rho,\sigma)$ which we will review soon. In the subsequent sections we  discuss how to probe soft limits while maintaining on-shell kinematics, as well as the construction of amplitudes from the above two criteria.

\subsection{Soft Momentum Shifts}
\label{sec:soft_shift}

Our analysis makes heavy use of the soft momentum shift proposed in~\cite{Cheung:2015ota}.  This deformation maintains total momentum conservation and on-shell conditions while probing the soft limits of external particles.   In \cite{Cheung:2015ota} these momentum shifts were used to construct new recursion relations for scattering amplitudes in EFTs.  However, here we need them as just a tool for probing the kinematics of scattering amplitudes.

The original soft momentum shift is applicable only when there are more than $d+1$ external legs in $d$ spacetime dimensions. 
In order to probe the full EFT space, we develop a number of simple variations on the soft momentum shift.
Although it seems to be a technical obstruction, we will see that the applicability has a one-to-one correspondence to the non-trivial soft limits in Sec.~\ref{sec:theory_space}.
We now discuss each momentum shift, whose properties are summarized in Table~\ref{tab:shifts}.

\begin{table}[t]
	\begin{center}
		\begin{tabular}{c|c|c|c|}
			{\bf Soft Shift} & {\bf Applicability}  & {\bf \# of Soft Limits} & {\bf Bound}   \\  \hline
			All-Line & $n>d+1$  & $n$ & $\frac{\rho-1}{\sigma-1} \geq \frac{v}{v-2}$ \\
			All-But-One-Line & $n>4$  &  $n-1$ &$\frac{\rho-2}{\sigma-2} \geq \frac{v-1}{v-2}$ \\ 
			All-But-Two-Line & $n\geq3$ and $d\geq 4$  & $n-2$ & $ \rho \geq \sigma  - \frac{2}{v-2}$ \\ 
		\end{tabular}
	\end{center}
	\caption{The first and second columns list soft momentum shifts and the conditions under which they can be applied to an amplitude with $n$ legs to probe its soft limits.  The third column lists the number of soft limits that are accessible by each soft shift when these criteria are satisfied. The fourth column lists the resulting constraints on EFTs with fixed $(\rho,\sigma, d,v)$ proved in Section~\ref{sec:leadingV}.  As discussed in text, these constraints are derived by applying each soft shift to the leading non-trivial amplitude, which is an amplitude with $n=v$ legs. }
	\label{tab:shifts}
\end{table}

\subsubsection*{All-Line Soft Shift}

We define the all-line soft shift by
\bea
p_i &\rightarrow&  p_i(1-za_i)\,, \qquad 1 \leq i \leq n\,,
\label{eq:all_shift}
\eea
where the shifted momenta are automatically on-shell but momentum conservation requires
\bea
\sum_{i=1}^n a_i p_i = 0\,. \label{eq:ai}
\eea
Since this constraint is a relation among the momenta, it may or may not be satisfied depending on the number of momenta $n$ relative to the space-time dimension $d$.  

There are two configurations of $a_i$ that are unphysical or not useful for probing the soft kinematic regimes of the amplitude.  First, one can rescale all the $a_i$ uniformly. This corresponds simply to a rescaling of the momentum deformation parameter $z$ and therefore not a new solution. 
Second, consider the case where the $a_i$ all equal.
This corresponds to a shift of the momentum of each leg by a constant times the momentum, which is also equivalent to a total rescaling of all the momenta. 
This class of momentum shifts does not probe any interesting kinematic regime of amplitudes provided the amplitude is a homogeneous function of momentum, which we assume here. 

The above two configurations can be viewed as the ``pure gauge'' configurations of $a_i$.
We can uniformly rescale or translate any solution of $a_i$ and the result is still a solution by Eq.~\eqref{eq:ai}. When counting degrees of freedom, the two pure gauge directions need to be excluded. Subtracting these two configurations, only $n-2$ degrees of freedom among the $a_i$ are of interest. The $d$ constraints of \Eq{eq:ai} then reduce these to $n-d-2$ independent variables.  Consequently, for $n \leq  d+1$, the momenta are linearly independent so there are either no solutions to \Eq{eq:ai} or the trivial configuration where all $a_i$ are equal.

Only for scattering amplitudes with sufficient numbers of external particles $n\ge d+2$ can we apply the soft shift in \Eq{eq:all_shift} with distinct $a_i$. 
In the marginal case $n=d+2$, the parameters $a_i$ are completely fixed up to rescaling and translation. There are residual degrees of freedom when $n>d+2$.
Note that the momentum conservation constraint in \Eq{eq:ai} implies that the $a_i$ are implicitly dependent on the $p_i$, constrained so they actually represent $n-d-2$ independent parameters.

Moreover, $z\rightarrow 1/a_i$ corresponds to taking the soft limit of particle $i$.  So for $n\ge d+2$ it is possible to apply an all-line soft shift that probes all the soft kinematic limits of the amplitude.



\subsubsection*{All-But-One-Line Soft Shift}

Similarly, we can define an all-but-one-line shift by
\bea
p_i &\rightarrow&  p_i(1-za_i)\,, \qquad 1\leq i \leq n-1\\
p_n &\rightarrow& p_n + zq_n\,,
\label{eq:all_but_one_shift}
\eea
where momentum conservation and the on-shell conditions imply that
\bea
q_n = \sum_{i=1}^{n-1} a_i p_i\,, &\quad& q_n^2 = q_n p_n=0\,.
\eea
Here we are shifting all the external legs, but in such a way that all-but-one of the soft limits can be accessed by taking $z\rightarrow 1/a_i$.

The all-but-one-line shift is defined by $n-1$ parameters $a_i$.  As before, the rescaling of $a_i$ and the case where all $a_i$ are equal correspond to a uniform rescaling of all the momenta, so only a subset of $n-3$ of these parameters are kinematically useful. 
Finally, the two on-shell conditions reduce these to $n-5$ independent variables, corresponding to distinct values of $a_i$.  

In summary, the all-but-one-line shift acts non-trivially on any amplitude with $n\ge 5$ legs in all dimensions, and which can probe $n-1$ soft limits.



\subsubsection*{All-But-Two-Line Soft Shift}

Lastly, we consider an all-but-two-line soft shift defined by
\bea
p_i &\rightarrow&  p_i(1-za_i)\,, \qquad  1\leq i \leq n-2\\
p_{n-1} &\rightarrow&  p_{n-1} + zq_{n-1}\,,\\
p_n &\rightarrow&  p_n + zq_n\,,
\label{eq:all_but_two_shift}
\eea
where momentum conservation and on-shell conditions imply
\bea
q_{n-1} +q_n = \sum_{i=1}^{n-2} a_i p_i\,, &\quad& q_{n-1}^2 = q_n^2 = q_{n-1} p_{n-1}= q_n p_n=0\,. 
\eea
Here we treat the $n-2$ parameters $a_i$ as free variables so that the two $d$-dimensional vectors $q_{n-1}$ and $q_n$ are constrained by the $d$ constraints from momentum conservation.
This corresponds to $d$ degrees of freedom subject to 4 constraints, leaving $d-4$ degrees of freedom in $q_{n-1}$ and $q_n$. Removing rescaling and translation as before, there are $n-4$ degrees of freedom in $a_i$. So the total number of independent variables are $(n-4)+(d-4)$.

In summary, for the general case $n\geq 5$, we find that the all-but-two-line soft shift acts nontrivially on any amplitude in $d\geq 3$ dimensions.
For the special case of 4pt, the all-but-two-line soft shift only works for $d \ge 4$ but not $d=3$.

\subsection{Soft Recursion Relations}
Next, we review the recursion relations for EFTs in \cite{Cheung:2015ota} 
(see also the generalization in \cite{Luo:2015tat}) which is a crucial tool for bounding the space of consistent EFTs.
To compute the $n$-pt amplitude, we first deform the momenta by any of the available soft shifts in \Sec{sec:soft_shift}. This promotes the amplitude $A_n$ into a function of $z$,
\be
A_n \rightarrow A_n(z)\,.
\ee
Then consider the contour integral
\begin{equation}
\oint \frac{dz}{z}\frac{A_n(z)}{F_n(z)} = 0\,,
\label{eq:Cauchy}
\end{equation}
where the denominator $F_n(z) = \prod_{i=1}^{n_s} (1-a_i z)^{\sigma}$.
The product in $F_n(z)$ runs from 1 to $n_s$, the number of external legs whose soft limit are accessible by the soft shift, given by the third column in Table~\ref{tab:shifts}. 
We can retrieve the original amplitude $A_n(0)$ by
choosing the contour as an infinitesimal circle around $z=0$. Cauchy theorem then relates the original amplitude as the (opposite) sum of all other residues.
The possible poles correspond to factorization (poles in $A_n(z)$), soft limit ($F_n(z)=0$), and the pole at infinity.
However, the integrand is designed such that $A_n(z)/F_n(z)$ has no pole in the soft limit $z=1/a_i$ since the amplitude vanishes as
\begin{equation}
A(z\rightarrow {1}/{a_i}) \,{\sim}\, (1-a_i z)^{\sigma},
\label{eq:soft_detail}
\end{equation}
as we define in Eq.~\eqref{eq:def_sigma}.
If there is no pole at infinity, the original amplitude is equal to the sum of residues from factorization channel. For each factorization channel $I$, there are two poles $z_{I\pm}$ corresponding to the roots of
\begin{equation}
P_I^2(z)=P_I^2 + 2 P_I \cdot Q_I z + Q_I^2 z^2 = 0\,,
\label{eq:poles}
\end{equation}
where $P_I(z) = P_I + z Q_i$ and where 
\begin{equation}
P_I = \sum_{i\in I} p_i \quad\textrm{and}\quad Q_I=-\sum_{i\in I}a_i p_i\,.
\label{eq:PQdef}
\end{equation}
By locality, each residue is a product of lower-point amplitudes. Applying Cauchy theorem then yields the recursion relation
\begin{equation}
A_n(0)= \sum_I \frac{1}{P_I^2}  \frac{A_{L}(z_{I-})\,A_{R}(z_{I-})}{(1-z_{I-}/z_{I+})F(z_{I-})} + (z_{I+} \leftrightarrow z_{I-})\,.
\label{eq:soft_recursion}
\end{equation}

The recursion relation above hinges on the absence of pole at infinity.
The large $z$ behaviors are $F_n(z) \sim z^{n_s \sigma}$ and $A_n(z)\sim z^{m}$ where $A_n$ has $m$ powers of momenta defined by Eq.~\eqref{eq:L}. The function $A_n(z)/F_n(z)$ vanishes at infinity provided $m<n_s \sigma$ which can be written as
\begin{equation}
\sigma > \frac{2+(n-2)\rho}{n_s}
\label{eq:constructibility}
\end{equation}
in terms of $\rho$ defined by Eq.~\eqref{eq:rho_def}.
Remember that depending on $n$ and $d$, each shift has its own applicability (see Table~\ref{tab:shifts}).
For exceptional theories, $\sigma=\rho+1$, we can use any of the three shifts in Sec.~\ref{sec:soft_shift} to construct the amplitude starting from 5pt. This implies 4pt amplitudes dictate all other amplitudes. 
For theories on the non-trivial line $\sigma=\rho$, all-line and all-but-one-line soft shift can construct amplitudes with $\sigma >1$ and $\sigma>2$ respectively. 
Note that all-but-two-line is no longer applicable on this line.
According to Table~\ref{tab:shifts}, theories with $\sigma=\rho=2$ like the general Galieon need the 4pt to $(d+1)$pt scattering amplitudes as seeds for the recursion relation.

\subsubsection*{Example: Six point amplitude in NLSM}
\label{sec:recursion_example}
As an illustration of these recursion relations, consider the 6pt amplitude in NLSM.
We use the all-but-one line soft shift so that our results apply in general dimensions.
This momentum shift is applicable in all exceptional theories for amplitudes above 4pt. The flavor-ordered 4pt amplitude reads
\begin{equation}
A_4=s_{12}+s_{23}\,.
\label{eq:NLSM_4pt}
\end{equation}
The recursion relation in Eq.~\eqref{eq:soft_recursion} can be rewritten as
\begin{equation}
\begin{split}
A_6(0) = -\sum_{I} \text{Res}_{z_{I\pm}}\left( \frac{A_L(z) A_R(z)}{z \, P^2_I(z) \, F(z)} \right)\,.
\end{split}
\label{eq:soft_recursion2}
\end{equation}
Note that we only probe soft limits of first five legs, so $F(z)=\prod_{i=1}^{5}f_i(z)$ where $f_i(z)=(1-a_i z)$.
For 6pt amplitude, the sub-amplitudes $A_L(z),A_R(z)$ are 4pt which have no poles.
Thus, we can use Cauchy theorem again term by term in the above equation
\begin{equation}
\begin{split}
A_6 & = \sum_{I} \left\lbrace \frac{A_L A_R}{P^2_I} +
\sum_{i} \underset{z={1/a_i}}{\text{Res}} \left(\frac{A_L(z) A_R(z)}{z P^2_I(z) F(z)} \right)\, \right\rbrace \\
& = \left[\frac{(s_{12}+s_{23})(s_{45}+s_{56})}{P_{123}^2} + \dots\right]+
\sum_{i,I} \underset{z={1/a_i}}{\text{Res}} \left(\frac{A_L(z) A_R(z)}{z P^2_I(z) F(z)} \right)\,,
\end{split}
\label{eq:soft_recursion3}
\end{equation}
where the first term is the residue at $z=0$, the second term sums over the residues from $F(z)=0$ which corresponds to the soft limits, and ellipses denote cyclic permutations. We will identify the second term as the contact term in the amplitude.
\begin{equation}
\begin{split}
A_{6,{\rm contact}} =
\sum_{i,I} \underset{z={1/a_i}}{\text{Res}}\left(\frac{A_L(z) A_R(z)}{z P^2_I(z) F(z)} \right)\,.
\end{split}
\label{eq:soft_recursion4}
\end{equation}

For the flavor-ordered 6pt amplitude there are three factorization channels 
corresponding to when $P_{123}$, $P_{234}$, and $P_{345}$ go on-shell.
The above contact term can be decomposed into
\begin{equation}
A_{6,{\rm contact}}= A_{6,{\rm contact}}^{(123)}+A_{6,{\rm contact}}^{(234)}+A_{6,{\rm contact}}^{(345)}\,.
\end{equation}
Considering the first term, we can plug Eq.~\eqref{eq:NLSM_4pt} into Eq.~\eqref{eq:soft_recursion4}, yielding
\begin{align}
A_{6}^{(123)} &=
-\left( \frac{\hat{s}_{45}+\hat{s}_{56}}{f_2 f_3 f_4 f_5}\right)\Big|_{z=1/a_1}
-\left( \frac{\hat{s}_{45}+\hat{s}_{56}}{f_1 f_2 f_4 f_5}\right)\Big|_{z=1/a_3}
-\left( \frac{\hat{s}_{12}+\hat{s}_{23}}{f_1 f_2 f_3 f_5}\right)\Big|_{z=1/a_4}\,.
\label{NLSMterm}
\end{align}
Here $\hat{s}_{ij}$ is the Mandelstam variable evaluated at shifted kinematics. 
Note that one of the sub-amplitudes cancels the propagator on the soft limit. For example, $P^2_{123}(1/a_1)=\hat{s}_{23}=A_L(1/a_1)$.
The residue at $z=1/a_1$ only shows up in $A_{6}^{(123)}$ and $A_{6}^{(234)}$. Combining the two yields
\begin{align}
-\left( \frac{\hat{s}_{23}+\hat{s}_{34}+\hat{s}_{45}+\hat{s}_{56}}{f_2 f_3 f_4 f_5} \right) \Big|_{z=1/a_1}
=\underset{z={1/a_1}}{\text{Res}}\left( \frac{\hat{s}_{12}+\hat{s}_{23}+\hat{s}_{34}+\hat{s}_{45}+\hat{s}_{56}+\hat{s}_{61}}{z f_1 f_2 f_3 f_4 f_5} \right),
\label{NLSMterm2}
\end{align}
where we include $\hat{s}_{12}+\hat{s}_{61}$ in the numerator in the right-hand side since they vanish at $z=1/a_1$.
All residues at $z=1/a_i$ can be combine into such form. 
Summing all of such gives
\begin{align}
\sum_{i=1}^{5} \underset{z={1/a_i}}{\text{Res}}\left( \frac{\hat{s}_{12}+\hat{s}_{23}+\hat{s}_{34}+\hat{s}_{45}+\hat{s}_{56}+\hat{s}_{61}}{z f_1 f_2 f_3 f_4 f_5} \right) = -(s_{12}+s_{23}+s_{34}+s_{45}+s_{56}+s_{61}),
\label{NLSMterm3}
\end{align}
where we use Cauchy theorem again to recast the sum into residue at the origin.
Combining the non-contact terms, the final answer is
\begin{equation}
A_6 = \left[\frac{(s_{12}+s_{23})(s_{45}+s_{56})}{P_{123}^2} + \dots\right]  - (s_{12}+\dots)\,,
\end{equation}
where ellipses again denote cyclic permutations. The above expression is the same one obtained via Feynman diagrams.

\section{Bounding Effective Field Theory Space}
\label{sec:theory_space}

With an arsenal of momentum shifts and on-shell recursion relations, we are now ready to ascertain the allowed parameter space of EFTs.
The aim of this section is to study the parameter space of EFTs as a function of $(\rho,\sigma, d,v)$ and determine regions of theory space which are inconsistent with locality and Lorentz invariance.  
To exclude swaths of EFT parameter space, we will consider several consistency checks.  The first will be a study of the soft limit of the leading interaction vertex of the EFT.  The second will be a study of the locality properties of higher point amplitudes.

\subsection{Soft Limit of the Leading Interaction}
\label{sec:leadingV}

Consider an EFT with the fixed $(\rho,\sigma, d,v)$.  All amplitudes in this EFT have soft degree $\sigma$ by assumption, including the leading non-vanishing amplitude $A_v$, where $v$ is the valency of the lowest point interaction. Since $A_v$ is comprised of a single vertex it has no factorization channels and is simply a polynomial function of the momenta. Given the definition of $\rho$ in \Eq{eq:rho_def}, this function contains $\rho (v-2)+2$ powers of momentum.

To begin, consider a soft momentum shifts in \Sec{sec:soft_shift} applied to $A_v$, lifting it to a complex function of $z$, so $A_v \rightarrow A_v(z)$.  Since $A_v$ is a contact amplitude, $A_v(z)$ is simply a polynomial in $z$. The degree of this polynomial is fixed by the mass dimension $\rho(v-2)+2$, since each momentum in the shift is linear in $z$.

At the same time, the vanishing soft limit corresponds to zeros of this polynomial. In particular, if $v_s$ is the number external legs whose soft limits can be probed by the soft momentum shift, then the total number of zeros are $v_s \sigma$ according to Eq.~\eqref{eq:def_sigma}. Comparing the degree of polynomial with the number of zeros yields
\begin{equation}
\rho \ge \frac{v_s\sigma-2}{v-2}\,.
\label{eq:contact_bound_2}
\end{equation}
Therefore, the most stringent bound on $\rho$ requires the maximal $v_s$.
Crucially, this depends on $v$ and $d$ as shown in \Sec{sec:soft_shift} and so does the bound on $\rho$.
These bounds are summarized in the fourth column of Table~\ref{tab:shifts}.

Altogether these bounds place a lower bound on $\rho$ as a function of $\sigma$ and $v$ which excludes almost all possible EFTs with non-trivial soft limits.  To explain these constraints, let us consider each of these bounds as a function of the leading interaction valency $v$ relative to the space-time dimension $d$.   Throughout, we assume space-time dimension $d\geq 4$.

The most general possible bounds arise from the all-but-two-line shift.  As we are concerned with scalar theories, the lowest possible valency of the leading interaction is $v=4$.  From Table~\ref{tab:shifts}, the bound is weakest---that is, places the smallest lower bound on $\rho$---for $v=4$ and becomes stronger as $v$ grows.  So conservatively, we can evaluate the all-but-two-line shift constraint from Table~\ref{tab:shifts} for $v=4$ to obtain a universal and remarkably stringent bound of
\bea
\rho &\geq& \sigma -1\,.
\label{eq:exceptional}
\eea
Notably, this bound is exactly saturated by the exceptional theories discussed in \cite{Cheung:2014dqa}, corresponding to the NLSM $(\rho,\sigma)=(0,1)$, DBI theory $(\rho,\sigma)=(1,2)$, and special Galileon $(\rho,\sigma)=(2,3)$.  Unsurprisingly, this result verifies that there are no theories with $\rho = 0,1,2$ with soft limits that are super-enhanced beyond these exceptional theories.  This is expected because these exceptional theories each have a single coupling constant and are thus already so constrained by soft limits that they have no additional free parameters.  Demanding a super-enhanced soft limit will over-constrain these theories, so no EFT exists with such properties.  Less obvious is the statement that for general $\rho$---including rational but non-integer values---there are no theories with soft limits enhanced beyond the exceptional line defined by \Eq{eq:exceptional}. Note that the proof here uses all-but-two-line shift which is valid only in $d\ge 4$. The same conclusion holds in $d=3$, which we will revisit in the end.

For the all-line and all-but-one-line shifts we obtain more stringent constraints which are applicable only in specific ranges for $v$ and $d$.
First, consider the constraint in Table~\ref{tab:shifts} from the all-line shift, which is applicable only when the valency $v$ of the leading interaction is greater than $d+1$.  The resulting bound on $\rho$ is a line that intersects the point at $(\rho,\sigma)=(1,1)$, which describes a derivatively coupled theory of a single NGB, called sometimes $P(X)$ theory (see Appendix~\ref{app:sumoftheories}).  The slope of the boundary is $v/(v-2) > 1$ so it is steeper than the $\rho = \sigma$ line that delineates the boundary between theories with trivial versus non-trivial soft limits.  Since $\sigma$ is a positive integer, we can exclude all EFTs with non-trivial soft limits for which $v>d+1$.
This result is consistent with the properties of known EFTs.  In particular, the Galileon theory is known to have interaction vertices up to $v=d+1$ valency but not higher.

Second, consider the constraint in Table~\ref{tab:shifts} from the all-but-one-line shift, which is applicable only when $4<v\leq d+1$.  Here the resulting bound intersects the Galileon theory at $(\rho,\sigma)=(2,2)$ with a slope of $(v-1)/(v-2)>1$, which is again steeper than $\rho=\sigma$.  Hence, this bound eliminates all EFTs with non-trivial soft limit $\sigma>2$ and $v>4$. 
The only allowed possibilities are then $(\rho,\sigma)=(2,2)$, which is consistent with the known Galileon theory,
or $\sigma=1$ with $\rho\geq {(v-3)}/{(v-2)}$ which is saturated by WZW theory. We will discuss the allowed region in depth in \Sec{sec:enumeration}.

The above bounds significantly simplifies the numerical search of possible theories. For a given dimension $d$, we only need to search leading amplitudes up to $v=d+1$. The inverse question is, given the leading valency $v$, what are the upper bound on spacetime dimension that we do not expect to find new non-trivial amplitudes?

The answer is given by a simple statement in kinematics.
For example, the 4pt kinematics in any $d\ge 3$, effectively lies in a three-dimensional subspace. This is easily seen in center of mass frame, where the four spatial momenta lie in a plane. The generalization to high dimension is straightforward: the $v$-pt kinematics in $d \ge v-1$ dimension only lives in a $(v-1)$-dimensional subspacetime.
If this is true, we can always take the soft limits within this $(v-1)$-dimensional subspacetime.
It implies the enhanced soft limit at $v$-pt in $d \ge v-1$ dimension must be present in $d=v-1$ already.
The numerical search up to $d=v-1$ can saturate all non-trivial amplitudes at arbitrarily higher dimension, which significantly reduces the space of possible theories that need to be checked.

The proof is analogous to 4pt. First consider the center of mass frame of the first two particles whose momenta are chosen as
\begin{equation}
\begin{split}
p_1 &=\frac{E_{\rm CM}}{2}(1,1,0,\dotsb ,0)\,, \\
p_2 &=\frac{E_{\rm CM}}{2}(1,-1,0,\dotsb ,0)\,.
\end{split}
\end{equation}
Next, due to total momentum conservation, only $v-3$ momenta of the remaining $v-2$ particles are independent. Using spatial rotations (or the standard Gram-Schmidt decomposition), we can choose a basis where these $v-3$ momenta lie in a $(v-3)$-dimensional subspace.
Together with the spatial part $p_{1,2}$, all spatial momenta can be chosen to reside in the first $v-2$ spatial components
\begin{equation}
\begin{split}
p_i =(E_i,p_{i1},\dotsb,p_{i,v-2},0,\dotsb ,0),\quad \forall\, i=\lbrace 3,\dotsb,v \rbrace\,.
\end{split}
\end{equation}
Combining with the temporal component, we find the $v$-pt kinematics only lives in a $(v-1)$-dimensional sub-spacetime as we claimed.

Let us come back to the case of $d=3$. First, the same bound from all-but-one-line and all-line shifts applies for $v\ge 5$ and $v\ge 6$ respectively. So we only need to consider 4pt case in $d=3$. Although we cannot use momentum shifts to prove Eq.~\eqref{eq:exceptional}, the 4pt kinematics always live in a three-dimensional subspace. Therefore, the 4pt kinematics should still satisfy $\rho =\sigma+1$ as in higher dimensions, which can be verified explicitly. So all the bounds are the same for $d=3$.

In summary, the leading valency $v$ of EFTs with an enhanced soft limit must satisfy
\begin{equation}
v\leq d+1\,,
\end{equation}
while the enhanced soft limit should be present in
\begin{equation}
d = v-1\,.
\end{equation}
These imply that for the numerical search of the non-trivial leading amplitudes, we can focus on the line of $v=d+1$.
Moreover, if $v>4$, then the soft degree and power counting parameters are bounded by
\begin{align}
\sigma=1 \textrm{ or } 2\,, \quad & \textrm{ and } \quad \rho\geq {(v-3)}/{(v-2)}\,.
\end{align}

\subsection{Locality of Higher Point Amplitudes}
\label{sec:locality}
The bounds derived in the previous section imply that the soft degree of an EFT cannot be exceed those of the exceptional EFTs.  Nevertheless, these constraints still permit an infinite band in EFT space between the exceptional line $\rho=\sigma-1$ and non-trivial line $\rho=\sigma$, as shown in \Fig{fig:EFT_map}.
While we can constructively identify the known theories with $\sigma=1,2,3$,
there is a priori no restriction on EFTs of arbitrarily high soft degree beyond $\sigma>3$, which we dub ``super-enhanced'' soft behavior.  However, in this section we show how EFTs with such super-enhanced soft behavior are impossible.

As discussed in the previous section, an exceptional EFT must have a valency $v=4$ for the leading interactions.  Without loss of generality, the corresponding 4pt contact amplitude takes the form
\begin{equation}
A_{4}= \sum_{b=0}^{\rho+1} \lambda_{b}\,s_{13}^{b}\, s_{12}^{\rho+1-b},
\label{eq:4pt_ansatz}
\end{equation}
where $\lambda_{b}$ are coupling constants.
From \Eq{eq:4pt_ansatz} we see that the soft degree is $\sigma=\rho+1$ but can in principle be arbitrarily large. Hence, there is of yet no obvious obstruction to a  theory with arbitrary high soft degrees.

To exclude such theories, we exploit the fact that exceptional theories are on-shell constructible \cite{Cheung:2015ota}.
Furthermore, in the previous section we showed that for $\sigma >2$, the only contact amplitudes consistent with non-trivial soft behavior enter at 4pt.  Altogether, this implies that all higher point amplitudes are fixed in terms of the 4pt amplitudes in Eq.~\eqref{eq:4pt_ansatz} via on-shell recursion.   Self-consistency then requires that the resulting higher point amplitude be independent of the precise way in which recursion is applied.  Concretely, the recursion relation should produce scattering amplitudes which are independent of the specific momentum shift employed.  For soft recursion relations, this means that the intermediate and unphysical momentum shift parameters $a_i$ should cancel in the final expression, since the physical amplitude should only depend on Mandelstam variables. As shown in the example in \Sec{sec:recursion_example}, such a cancellation is highly non-trivial.  In the following, we study this cancellation and use it to derive a no-go theorem for the existence of super-enhanced theories.

Our approach mirrors the so-called ``four-particle test'' of \cite{Benincasa:2007xk} (and see also \cite{McGady:2013sga}), where the consistency of higher spin theories was similarly studied via on-shell recursion.  There it was shown that for theories of massless particles of spin greater than two, recursion relations yield different answers depending on the momentum shift used.  This failure of recursion relations indicates an underlying tension between locality, factorization, and gauge invariance in the underlying theory.  The same logic can be applied here: if soft recursion relations yield dependence on unphysical parameters in the final answer, then it is impossible to construct higher point amplitudes which are simultaneously local with the correct soft and factorization properties.

Since the details of the proof are rather technical, readers can skip the following and move to Sec.~\ref{sec:enumeration} if they are uninterested in the details.  However, our final results from this analysis are that:
\begin{itemize}
\item All EFTs with non-trivial soft behavior have $\rho <3$.  This claim is independence of flavor structure, and applies for single or multiple scalar EFTs.

\item The NLSM is the unique EFT with flavor-ordered amplitudes that exhibit exceptional soft behavior, $\sigma = \rho+1$.

\end{itemize}
We find the locality test imposes a stringent bound on the theory space of EFTs, as shown in Figure~\ref{fig:EFT_map}. Galileon theories live on the boundary of the allowed region.

\subsubsection*{Details of the Proof}

\begin{figure}[t]
	
	\begin{center}
		\includegraphics[width=.6\textwidth, trim =6cm 12cm 6cm 5cm, clip]{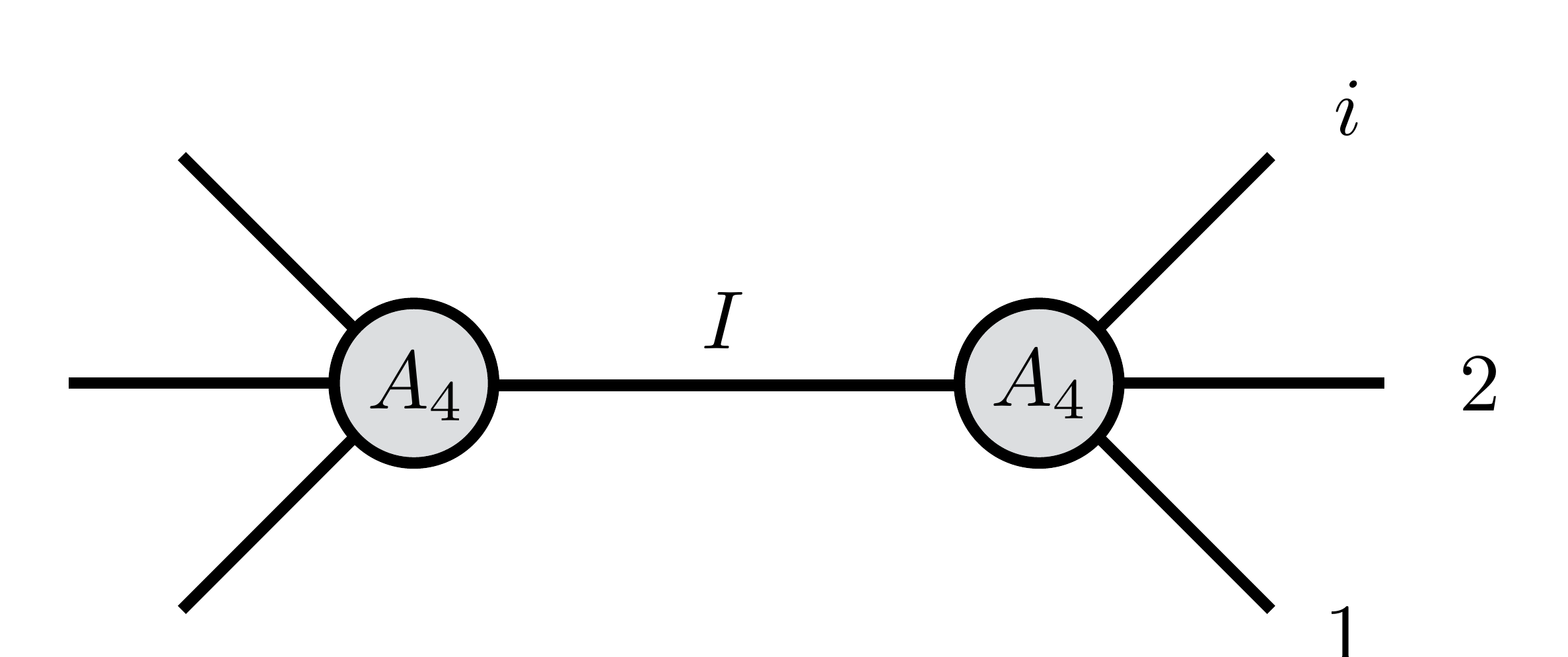}
	\end{center}
	\caption{Factorization channel with spurious pole $a_{12}$.}
	\label{fig:fact} 
\end{figure}

We diagnose the self-consistency of super-enhanced soft behavior by analyzing the 6pt amplitude, in analogy of the 4pt test in higher spin gauge theory.  Specifically, we
consider the 6pt kinematics in $d=3$ where we are allowed to apply all-line soft shift.
For higher dimensional theories, we can always take a special 6pt kinematics restricted to $d=3$.
One might worry that the 6pt amplitude vanishes in this limit and thus trivializes the test.
However, the non-trivial soft limits wiht $\rho>2$ fix all amplitudes from 4pt amplitudes via the recursion relations.
As we discussed in Section~\ref{sec:leadingV}, the 4pt kinematics in $d=3$ is already generic.
We will see the spurious pole cancellation put constraints on the 4pt coupling constants.
If the only consistent coupling constants are zero in the $d=3$ special kinematics, then the 6pt amplitudes, which are given by the recursion, must be trivial even in generic kinematics.
Therefore, the proof here applies to general $d\ge 3$.

Let us consider the 6pt amplitude obtained from recursion relations.
As shown in Eq.~\eqref{eq:soft_recursion3}, it can be decomposed into factorization terms (comprised of two 4pt vertices and a propagator) and the contact term (comprised of one 6pt vertex). The example presented in Eq.~\eqref{eq:soft_recursion3} is for the NLSM, but this decomposition is generally applicable.

First, we see that the factorization terms are manifestly independent of the shift parameters $a_i$.  Hence, these cannot contain any spurious dependence on the momentum shift so we can ignore them.  On the other hand, the contact term reads
\begin{equation}
\begin{split}
A_{6,{\rm contact}} =
\sum_{i,I} \underset{z={1/a_i}}{\text{Res}} \left(\frac{A_L(z) A_R(z)}{z P^2_I(z) F(z)} \right)\,,
\end{split}
\label{eq:recursion_contact}
\end{equation}
which can in principle depend on $a_i$, yielding an inconsistency.  Conversely, consistency implies that \Eq{eq:recursion_contact} is $a_i$ independent, so all spurious poles in these parameters must cancel.  Here unphysical poles in $a_i$ can only appear in the denominator of \Eq{eq:recursion_contact} because $A_{L,R}$ are 4pt amplitudes which are local functions of momenta, and thus local functions of $a_i$.

Let us determine what kind of spurious poles can arise from the above equation.
Recall that $F(z)=\prod_{j=1}^{6}f_i^{\sigma}(z)$, where $f_j(z)=1-a_j z$, is the product of rescaling factors. Furthermore, observe that the rescaling factor of leg $j$ evaluated at $z=1/a_i$ is proportional to $(a_i -a_j)$, which induces a spurious pole. 
In general, the shifted propagator can also contain a similar form of spurious pole: for example, $P^2_{123}(1/a_2)=f_1 f_3 s_{13}$ is proportional to $(a_2-a_1)(a_2-a_3)$.  

In what follows we analyze the unphysical pole at $a_1 \rightarrow a_2$ and show that the criterion that this singularity cancels in the final amplitude imposes a constraint on allowed EFTs. 
Here it was important that we can take the all-line soft shift in $d=3$ at 6pt, so it is possible to send $a_1 \rightarrow a_2$ while keeping all other $a_i$ distinct. 
Taking residue at $z=1/a_{2}$ is then reminiscent of a double soft limit, where leg $2$ is exactly soft, $p_2(1/a_{2})=0$, and leg $1$ approaches soft $p_1(1/a_{2})\sim (a_1-a_2)p_1$ as $a_1 \rightarrow a_2$.  
As explained in the previous paragraph, the spurious pole in $a_1-a_2$ only appears when taking  the residue at $z=1/a_1$ or $z=1/a_2$.   Legs 1 and 2 either appear on opposite sides of the factorization channel, or the same side, which we now consider in turn.

If legs 1 and 2 are on different sides of factorization channel, we can always parametrize the 4pt amplitudes as
\begin{equation}
\begin{split}
A_{L}(z) & = \sum_{b=0}^{\rho+1} \lambda_{b}\,\hat{s}_{1i}^{b}\, \hat{s}_{1j}^{\rho+1-b} \propto f_1^{\rho+1}(z)\,, \\
A_{R}(z) & = \sum_{b=0}^{\rho+1} \lambda_{b}\,\hat{s}_{2k}^{b}\, \hat{s}_{2l}^{\rho+1-b} \propto f_2^{\rho+1}(z)\,,
\end{split}
\label{eq:4pt_ansatz_split}
\end{equation}
where $i,j,k,l$ label the on-shell legs in the amplitude other than legs 1 and 2.
Recall that hatted Mandelstam variables are evaluated at shifted kinematics.
Meanwhile, the internal propagator, $P^2_I(z)$ will never be singular as $a_1 \rightarrow a_2$, since the double soft limit does not yield a singularity from the propagator in this channel.  Since $F(z) \propto f_1(z)^\sigma f_2(z)^\sigma$, \Eq{eq:4pt_ansatz_split} implies that the overall scaling of the contact factorization term is $(f_1 f_2)^{\Delta}$, where for later convenience we define
\bea
\Delta= \rho+1 -\sigma\,.
\label{eq:Deltadef}
\eea
Here $\Delta =0$ for exceptional EFTs, while $\Delta=1$ for EFTs with non-trivial behavior.  Meanwhile, $\Delta >1$ EFTs have trivial soft behavior that is guaranteed simply by large numbers of derivatives, and $\Delta<0$ is forbidden by the arguments from the contact amplitude in the previous section.  Putting this all together, since $\Delta$ is strictly non-negative, these terms can never produce a spurious pole as $a_1 \rightarrow a_2$.

Therefore, the spurious pole only appears when legs 1 and 2 are on the same side of the factorization channel. Namely, we only need to consider factorization channel $I=123,124,125,126$ as shown in Figure~\ref{fig:fact}. In this case it is convenient to parametrize the 4pt amplitude
\begin{equation}
A_{4} (z)= \sum_{b=0}^{\rho+1} \lambda_{b}\,\hat{s}_{1i}^{b}\, \hat{s}_{12}^{\rho+1-b} \propto f_1^{\rho+1}(z) f_2^{\rho+1-b}(z)\,,
\label{eq:4pt_ansatz_12}
\end{equation}
without loss of generality and where $i=3,4,5,6$. 
This is chosen so that the 4pt amplitude carries a factor of $f_1^{\rho+1}(z)$ that will overpower the $f_1^\sigma(z)$ factor in the denominator of the recursion.  Thus, we find that spurious poles in  $a_1 \rightarrow a_2$ are localized to the residue from $f_2$, {\it i.e.}~the residue at $z=1/a_2$ in four factorization channels $I=123,124,125,126$.

Consider the factorization $I=12i$. We now combine the parameterization of the 4pt amplitude in Eq.~\eqref{eq:4pt_ansatz_12}, together with the recursion relation in \Eq{eq:recursion_contact} to localize the spurious pole in $a_1 \rightarrow a_2$.
We need to take the residue at $z=1/a_2$ from
\begin{equation}
\begin{split}
\frac{A_L(z) A_R(z)}{z P^2_{I}(z) F(z)} = \sum_{b=0}^{\rho+1} s_{12}^{\rho+1-b} \,  \left(
\frac{A_L(z)\,\lambda_{i,b}s_{1\hat{i}}^{b} f^{\Delta}_1(z)}
{z P^2_{I}(z) F_{3456}(z) f^{b-\Delta}_2(z)}
\right),
\end{split}
\label{eq:spurious_1}
\end{equation}
where $s_{1\hat{i}}=2p_1\cdot p_i(z)$, $F_{3456}(z)=(f_3 f_4 f_5 f_6)^{\sigma}$,  and $\Delta$ is defined as in \Eq{eq:Deltadef}.  
Here we have kept the dependence of coupling constant on $i$. The pole at $z=1/a_2$ in the above equation is generally not a simple pole. The residue is then obtained through taking derivatives.
However, the inverse propagator at $z=1/a_2$ contains spurious pole but not its derivative
\begin{equation}
\begin{split}
P^{2}_{I}(1/a_2) &= f_1(z) s_{1\hat{i}} \Big \vert_{z=1/a_2}\\
\frac{dP^{2}_{I}}{dz}(1/a_2) 
&\xrightarrow{a_1\rightarrow a_2} -a_2( s_{1\hat{i}}+ s_{2\hat{i}})\Big \vert_{z=1/a_2}
\end{split}
\end{equation}
Therefore, the leading spurious pole in the residue occurs when all the derives act on $P^2_I(z)$ but not on the numerators.
The highest number of derivatives needed to take happens for the largest $b$ where $\lambda_b \neq 0$ in Eq.~\eqref{eq:spurious_1}.

Now we combine everything together. First take the residue from Eq.~\eqref{eq:spurious_1} and only keep the leading spurious term from $b_{\rm max}$. Then, sum over factorization channels $I=123,124,125,126$. Finally we find
\begin{equation}
\begin{split}
\frac{1}{ (a_1-a_2)^{b_{\rm max}-2\Delta}} \times \left[
\sum_{i=3}^{6}  A_L(1/a_2)\,\lambda_{i,b_{\rm max}} \, s_{1\hat{i}}^{\Delta} (s_{1\hat{i}}+ s_{2\hat{i}})^{b_{\rm max}-\Delta-1}
\right]_{z=1/a_2}\,,
\end{split}
\label{eq:leading_spurious_normal}
\end{equation}
where we drop the irrelevant proportional constant. The spurious pole cancellation implies the numerator in the square bracket must vanish whenever the spurious pole forms, {\it i.e.}, $b_{\rm max}>2\Delta$.

In principle, there are several ways the above numerator can vanish. The most naive way is to forbid coupling constants whenever the spurious pole appears. 
The cancellation could also happen in the state sum in the multiple scalar case. The second possibility is to cancel the numerator in the summation of factorization channels. We only know the sufficient conditions for this to happen, which we will describe soon. But {\it a priori}, there could be accidental cancellations beyond our expectation and we have to check numerically for a given $b_{\rm max}$. Strictly speaking, this is a loophole since we cannot check arbitrary high $b_{\rm max}$ numerically. 
However, we can localize the spurious pole to one single factorization using the so-called ``bonus'' relation. In such a case, the spurious pole cannot appear at all. We can close the loophole by combining numerical checks to sufficiently high $b_{\rm max}$ and after that using the proof via bonus relations.
This proof using bonus relations is presented in Appendix~\ref{app:proofbonus}.
Hence, we will assume no such accidental cancellation in what follows.

In the following, we will first discuss sufficient conditions for the spurious pole cancellation, which are satisfied by all known EFTs.
These conditions are also necessary as supported by numerical checks and proofs from bonus relations.
We will then show bounds in single and multiple scalars in turn.

\subsubsection*{Locality Test in Known EFTs}
In the case of a single scalar, all the above constraints simplify dramatically since 
there is no state sum over flavors and the coupling constants $\lambda^{i,b}$ are universal.
Moreover, when $a_1 \rightarrow a_2$ and $z$ is evaluated at $1/a_2$, particle 1 and 2 are both soft.
The left sub-amplitude $A_{L}$ is then the universal 4pt amplitude of particle 3,4,5,6 which cannot be zero. We can furthermore factor out $A_L$ in Eq.~\eqref{eq:leading_spurious_normal}.

Consider exceptional theories which $\Delta=0$. Stripping off the universal $A_L$ and coupling constants in Eq.~\eqref{eq:leading_spurious_normal} yields the numerator
\begin{equation}
\begin{split}
\sum_{i=3}^{6} 
(s_{1\hat{i}}+ s_{2\hat{i}})^{b_{\rm max}-1}\,,
\end{split}
\label{eq:leading_spurious_normal_exceptional}
\end{equation}
which is evaluated at $z=1/a_2$. This has to vanish for $b_{\rm max}>0$ in generic kinematics. Recall that $z=1/a_2$ corresponds to the double soft limit on the first two legs. The rest of momenta $\hat{p}_{i}$ form a 4pt kinematics,
$\sum_{i=3}^{6}\hat{p}_{i}|_{z=1/a_2} =0$.
Therefore Eq.~\eqref{eq:leading_spurious_normal_exceptional} is satisfied if 
\begin{equation}
b_{\rm max}=2\,.
\end{equation}
We check numerically up to $b_{\rm max}=10$, above which are ruled out by the bonus relations in Appendix~\ref{app:proofbonus}.

DBI straightforwardly satisfies the constraint because $\rho=1$. On the other hand, the cancellation of spurious pole in special Galileon realizes in an interesting way
\begin{equation}
A_4 = s_{12}^3+s_{13}^3+s_{23}^3 =-3s_{12}^2s_{13}-3s_{12}s_{23}^2\,.
\end{equation}
Although the amplitude has terms $\sim s^3$, on-shell kinematics cancels the leading term and satisfies the locality constraint.

For theories with flavor-ordered amplitudes, e.g., NLSM, it is almost the same as single scalar except that we only sum over adjacent factorization channels and $\lambda_{b_{\rm max}}$ depends on the ordering. 
We cannot cancel spurious pole from $b_{\rm max}=2$ because global momentum conservation is no longer available when only adjacent factorization channels are summed. This can be checked numerically or be proven by the bonus relations in Appendix~\ref{app:proofbonus}.
However, the spurious pole for $b_{\rm max}=1$ can be canceled if
\begin{equation}
\lambda_{3,1}+\lambda_{6,1}=0\,.
\end{equation}
We can check explicitly that the cyclic 4pt amplitudes in the NLSM are
\begin{equation}
\begin{split}
A_4(1,2,3,I_{123}) & = -s_{13}\,, \\
A_4(6,1,2,I_{612}) & = s_{16}+s_{12}\,.
\end{split}
\end{equation}
So the coupling constants indeed have opposite sign and cancel the spurious pole.

For theories on non-trivial line, $\Delta=1$. There is an extra factor of $s_{1\hat{i}}$ that ruins all the previous cancellation. Therefore, we do not know any sufficient condition to cancel spurious pole in the sum. This constrains
\begin{equation}
b_{\rm max} \le 2\,,
\label{eq:bmax_suff_bound}
\end{equation}
as the same as exceptional theories. Again, we check numerically up to $b_{\rm max}=10$ and beyond which is ruled out by the bonus relations.

We point out there is an intriguing similarity between exceptional EFTs and YM and gravity. Here we find the locality in DBI and special Galileon hinges on global momentum conservation, and locality in NLSM relies on cancellation between adjacent channels. This is completely analogous to the mechanism of how gauge invariance is realized in soft theorems in YM and gravity~\cite{Weinberg:1965nx}. This could be a hint that these exceptional EFTs are closely related to YM and gravity.

\subsubsection*{Bounds on Single Scalar EFTs}

As discussed before, we can factor out coupling constants and the sub-amplitude $A_L$ in the case of single scalar. The locality test then demands $b_{\rm max} \le 2$.
On the other hand, any pair of $a_i,a_j$ could form a spurious pole. We can check spurious pole in $a_1-a_3$ from the parametrization of Eq.~\eqref{eq:4pt_ansatz_12}. The same bound applies if we replace $b$ with $\rho+1-b$. Combining the two bounds on $b$, we find
\begin{equation}
2 \ge b \ge \rho+1-2\,,
\label{eq:bound_b}
\end{equation}
which could be satisfied if $\rho \le 3 $. We find that $\rho$ cannot be arbitrary.

Moreover, we can discuss spurious pole in $a_2-a_3$ and parametrized the 4pt amplitude using any two of the Mandelstam variable $s_{23},s_{21},s_{31}$. The same bound $2 \ge b$ in Eq.~\eqref{eq:bound_b} still applies to the power of any Mandelstam variable in any parametrization.
From Eq.~\eqref{eq:bound_b}, the only permitted ansatz in $\rho=3$ is $A_4 \propto s^2_{13}s^2_{12}$.
This is not allowed in the basis where we replace $s_{13}$ with $-(s_{12}+s_{23})$.
We conclude that for any non-trivial theories with a single scalar, 
\begin{equation}
\rho \le 2\,, 
\end{equation}
which is saturated by Galileon theories.

We can also bound theories with flavor-ordered amplitudes. They are very similar to single scalar theory except that only adjacent factorization channels are included.
The spurious pole of $a_1-a_2$ only appears in channels $I=123,612$ and the spurious pole of $a_1-a_3$ only appears in $I=123$.
As discussed in the locality test of the NLSM, the cancellation of $a_1-a_2$ only works with $b_{\rm max} \le 1$ because we lose momentum conservation.
On the other hand, the spurious pole of $a_1-a_3$ only appears in $I=123$ and there is no cancellation. This demands $\rho+1-b_{\rm max}=0$ in the ansatz of Eq.~\eqref{eq:4pt_ansatz_12}. Combining both, we find 
\begin{equation}
\rho = 0
\end{equation}
for exceptional theories with stripped amplitudes. The 4pt stripped amplitude with $\rho=0$ is unique, which coincides with the NLSM one. As higher point amplitudes are uniquely specified by recursion, we conclude that NLSM is the \emph{unique} exceptional theory with flavor ordering.


\subsubsection*{Bounds on Multiple Scalar EFTs}

Next, let us consider the case of EFTs with multiple scalars.  As noted in earlier, some such theories admit flavor-ordered amplitudes, but this is not generic.
We consider the generic multi-scalar case without assuming flavor-ordering here.

There are two complications in the case of multiple scalars. First, the coupling constant $\lambda_{i,b}$ now depends on the scalar species. Second, we need to sum over all possible intermediate states in \Fig{fig:fact}.
The sub-amplitude $A_{L}$ is not universal and can no longer be factored out.

For example, consider the factorization channel $I=123$. The subamplitude ansatze are
\begin{equation}
\begin{split}
A_{4}(123I) &= \sum_{b=0}^{\rho+1} \lambda_{123I,b}\,s_{13}^{b}\, s_{12}^{\rho+1-b}\,, \\
A_{4}(456I) &= \sum_{b'=0}^{\rho+1} \lambda_{456I,b'}\,s_{45}^{b'}\, s_{46}^{\rho+1-b'}\,,
\end{split}
\label{eq:4pt_ansatz_multi}
\end{equation}
where $I$ labels an internal state.
The key observation is that the internal state dependence only affects the coupling constants. So in the recursion, the coupling constants will only appear in a particular form
\be
\lambda^{123}_{b,b'} \equiv \sum_{I} \lambda_{123I,b}\,\lambda_{456I,b'}\,,
\ee
where intermediate states $I$ are summed over.

Even without knowing individual coupling constants, it is sufficient to constraint the $\lambda^{123}_{b,b'}$, where we dub ``coupling constant square''.
If all of them are zero, then the 6pt amplitude must be trivial from recursion.
This implies the 8pt amplitude is zero because it factorizes into 4pt and 6pt ones. All the higher point amplitudes are then trivial by iterating this argument.
We will focus on the constraints on these coupling constant square in the following.


Plugging the ansatze in Eq.~\eqref{eq:4pt_ansatz_multi} into Eq.~\eqref{eq:leading_spurious_normal}, the spurious pole cancellation requires 
\begin{equation}
\begin{split}
\sum_{i=3}^{6} \sum_{b'=0}^{\rho+1}
\lambda^{12i}_{b_{\rm max},b'} \,
\left(s_{\hat{4}\hat{5}}^{b'}\, s_{\hat{4}\hat{6}}^{\rho+1-b'} \right)
s_{1\hat{i}}^{\Delta} (s_{1\hat{i}}+ s_{2\hat{i}})^{b_{\rm max}-\Delta-1} = 0
\end{split}
\label{eq:leading_spurious_multi3}
\end{equation}
for $b_{\rm max}>2\Delta$.
We can check numerically if there is any choice of $\lambda^{12i}_{b_{\rm max},b'}$ that can solve the above equation for generic kinematics for given $b_{\rm max}$ and $\rho$.
We do not find a numerical solution for $b_{\rm max}>2$ for both exceptional theories  and non-trivial theories, up to $\rho=9$.
The bonus relations in Appendix~\ref{app:proofbonus} further rule out any such solution with $\rho>9$.
This constrains the coupling constant square $\lambda^{12i}_{b,b'}$ to have $b \le 2$ for any $b'$.

Following the same the spurious pole analysis on any pair of $a_i-a_j$,
both indices of the coupling constant square $\lambda^{123}_{b,b'}$ are restricted to be less or equal to two, in any choice of Mandelstam variable basis.
We find the same bound as Eq.~\eqref{eq:bound_b}.
As before, the ansatz of $\rho=3$ is restricted to $s^2_{12}s^2_{13}$ which is ruled out when switching to the basis of $s^2_{12}(s_{12}+s_{23})^2$.
We conclude that the bounds on multi-scalar EFTs are identical to single scalar EFTs.

\section{Classification of Scalar EFTs}
\label{sec:enumeration}

In the previous sections we derived stringent exclusions on the $(\rho,\sigma,v,d)$ parameter space of EFTs.  However, these exclusions still allow for EFTs to exist in the range
\begin{equation}
d+1 \geq v \quad \textrm{ and } \quad 3> \rho\geq\frac{\sigma(v-1)-2}{v-2}\,. \label{con}
\end{equation}
In what follows, we explicitly enumerate all scalar EFTs with non-trivial soft behavior, as defined by the window in \Eq{eq:triviality}.  A priori, this would require scanning over values of $(\rho,\sigma,v,d)$ and numerically determining whether there exists an amplitudes ansatz consistent with these assumptions.  However, as shown earlier, for a given choice of $(\rho,\sigma,v)$ it is always sufficient to check for the existence of EFTs in $d=v-1$ dimensions, since no new theories can appear for $d>v-1$. Thus for a given $v$ we only have to check all possible $(\rho,\sigma)$ regions in $d=v-1$ dimensions. 

In this section we enumerate and classify all possible EFTs for $v=4,5,6$, which in turn exhausts all possible theories in $d=3,4,5$. Our analysis begins with $v=5$ and $v=6$ theories, checking $n=v$ amplitudes. The $v=4$ is special because the 4pt amplitude does not give any constraints since $\sigma=\rho+1$ from 4pt kinematics. In this case we have to proceed further and consider 6pt amplitudes. 

We distinguish between cases with permutation invariance among legs (corresponding to amplitudes of a single scalar) or cyclic invariance (corresponding to flavor-ordered amplitudes of multiple scalars).
Note that for a single scalar with $\rho=0$, the permutation invariant amplitudes ansatz vanishes identically because any Lagrangian of that form is just field redefinition of free scalar field theory.  However, for multiple scalars with flavor-ordering, there is a non-trivial amplitudes ansatz. 

\subsection{Low Valency}

In this subsection we enumerate scalar EFTs whose leading interactions are at low valency, corresponding to $v=4,5,6$.

\subsubsection*{Case 1: $v=5$}

We begin with the case of leading valency $v=5$.  Here the corresponding critical dimension is $d=4$, by which we mean that it is sufficient to scan for theories in $d=4$ dimensions to enumerate all possible EFTs.  Analyzing amplitudes in higher dimensionality is unnecessary simply because the kinematics of the $v=5$ amplitude are constrained to $d=4$ anyway.   

We only consider EFTs which have non-trivial soft behavior and are thus on-shell constructible, so $\sigma\geq \rho$.  Moreover, we restrict to the region defined in \Eq{con},
\begin{equation}
3>\rho\geq\frac{4\sigma-2}{3}\,,\label{con5}
\end{equation}
which is in principle still permitted from our previous arguments.
For $v=5$, the only possible allowed pairs of $(\rho,\sigma)$ compatible with (\ref{con5}) and non-triviality bound are $(\rho,\sigma)=(\frac23,1)$ and $(\rho,\sigma)=(2,2)$.

\begin{figure}[t]
	\begin{center}
		\includegraphics[height=0.55\textwidth]{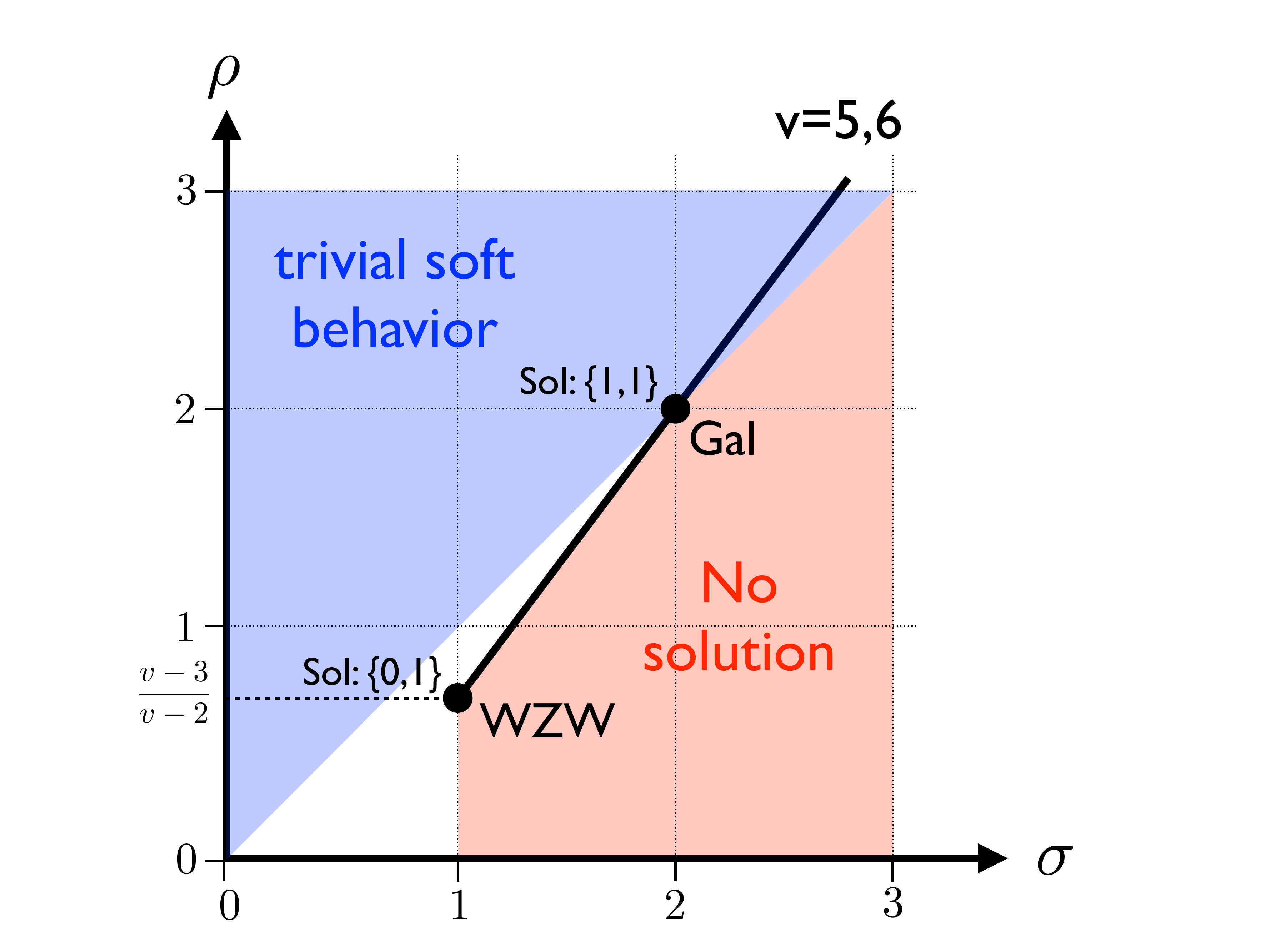}
	\end{center}
	\vspace{-5mm}
	\caption{
		Plot summarizing the numerical search of EFTs with $v=5,6$.
		The blue region denotes the same trivial region as in Figure \ref{fig:EFT_map}. 
		The red region has no solution numerically. The only two points with solutions are the $d$-dimensional WZW theory, $(\rho,\sigma)=(\frac{v-3}{v-2},1)$, and the Galileon $(\rho,\sigma)=(2,2)$. The label ``Sol:\{a,b\}'' denotes the number of solutions in permutation invariant and cyclic invariant amplitudes respectively.
		}
	\label{fig:num_checks}
\end{figure}

In Figure~\ref{fig:num_checks}, we use the symbol $\{a,b\}$ where $a$ denote the number of solutions in permutational invariant case and $b$ the number of solutions in cyclically invariant case.  We also performed checks for cases satisfying $\sigma\geq \rho$ and $\rho<3$ bounds but failing to meet Eq.~\eqref{con5}. There is no solution and the previous proof is confirmed.

We see from the diagram that there is one interesting 5pt cyclically ordered amplitude for $(\rho,\sigma)=(\frac23,1)$,
\begin{equation}
A_5^{(\frac23,1)} = \epsilon_{\mu\nu\alpha\beta} p_1^\mu p_2^\nu p_3^\alpha p_4^\beta\,,
\end{equation}
which arises precisely from the WZW term on the NLSM mentioned earlier. The presence of the Levi-Civita tensor implies that this solution exists only in $d=4$ and not other dimensions.

Another interesting solution appears for $(\rho,\sigma)=(2,2)$, and in $d=4$ can be compactly represented by
\begin{equation}
A_{5}^{(2,2)} = \left(\epsilon_{\mu\nu\alpha\beta} p_1^\mu p_2^\nu p_3^\alpha p_4^\beta\right)^2.
\end{equation}
In higher dimensions $d\ge4$ this amplitude takes the form
\begin{equation}
A_{5}^{(2,2)} =\delta^{\mu_1\ldots \mu_4}_{\nu_1\ldots\nu_4}p_{1\mu_1}\ldots p_{4\mu_4} p_1^{\nu_1}\ldots p_4^{\nu_4}\,,\label{gal5}
\end{equation}
which is equal to the Gram determinant since 
$\delta^{\mu_1\ldots \mu_n}_{\nu_1\ldots\nu_n} = \det (\delta^{\mu_i}_{\nu_j})_{i,j=1}^n$.
Such amplitudes are both cyclic and permutational invariant in all legs. This amplitude arises from the 5pt interaction of the Galileon theory, for both a single and multiple scalar fields (cf. Appendix~\ref{app:sumoftheories}), which exists in $d\geq4$. This exhausts all interesting cases for leading valency $v=5$.

\subsubsection*{Case 2: $v=6$}

For valency $v=6$, it is sufficient to study EFTs restricting to the critical dimension $d=5$ and the region in \Eq{con}, 
\begin{equation}
3>\rho\geq\frac{5\sigma-2}{4}\,.\label{con6}
\end{equation}
For $v=6$, the only non-trivial pairs $(\rho,\sigma)$ satisfying (\ref{con6}) are $(\rho,\sigma)=(\frac34,1)$ and $(\rho,\sigma)=(2,2)$. Indeed, there are two solutions for amplitudes, one for each point in the parametric space,
\begin{equation}
A_6^{(1,1)} =  \epsilon_{\mu\nu\alpha\beta\kappa} p_1^\mu p_2^\nu p_3^\alpha p_4^\beta p_5^\kappa\,,
\end{equation}
valid only in $d=5$ which corresponds to the WZW model. The other solution is the 6pt Galileon, written in $d=5$ as 
\begin{equation}
A_{6}^{(2,2)} = \left(\epsilon_{\mu\nu\alpha\beta\kappa} p_1^\mu p_2^\nu p_3^\alpha p_4^\beta p_5^\kappa\right)^2\,,
\end{equation}
but in general $d>4$ it takes the form (\ref{gal5}) with five momenta involved. 

\subsubsection*{Special Case: $v=4$}

\begin{figure}[t]
	\begin{center}
		\includegraphics[height=0.55\textwidth]{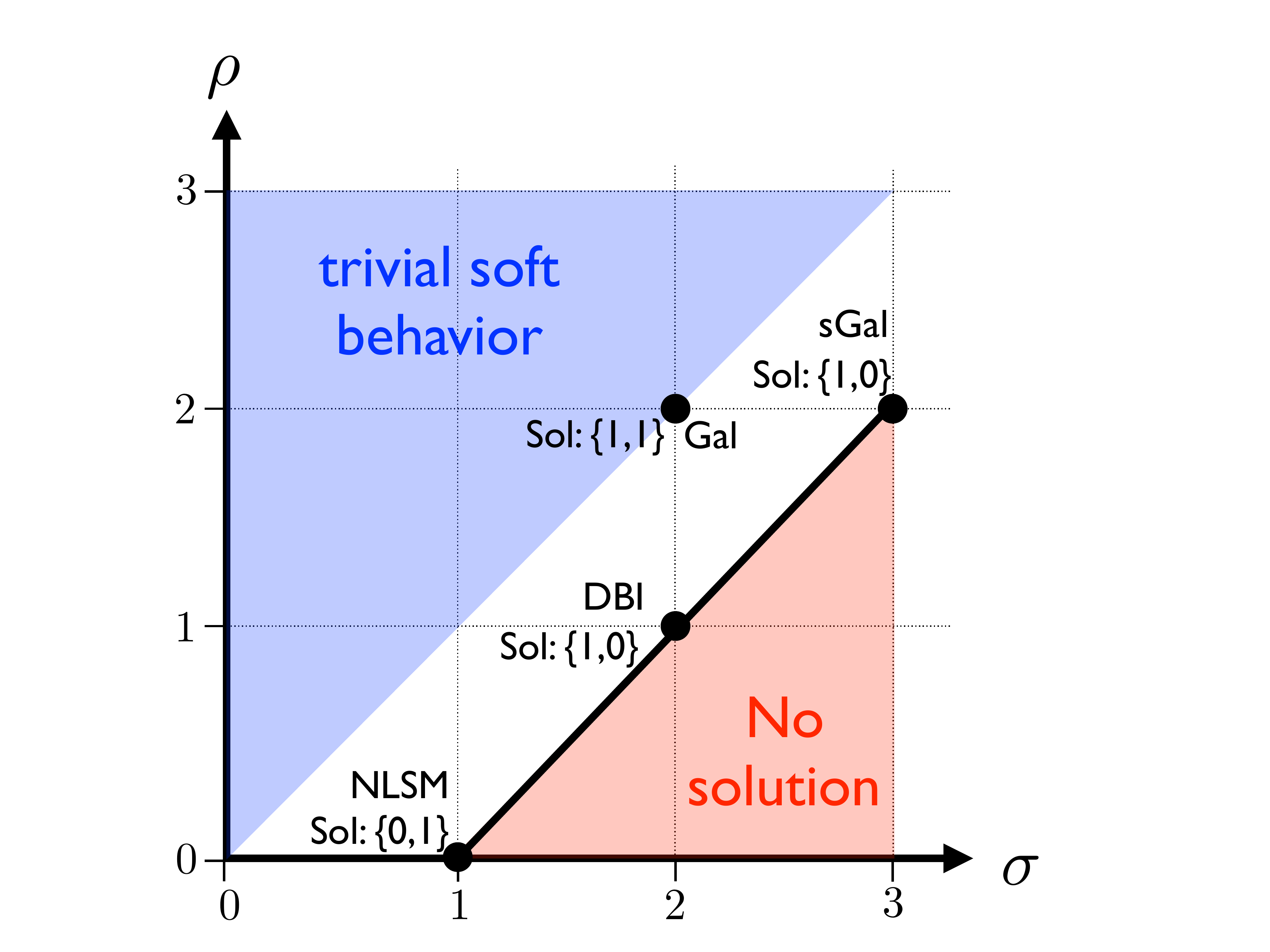}
	\end{center}
	\vspace{-5mm}
	\caption{
		Plot summarizing the numerical search of EFTs with $v=4$.
		The blue region denotes the same trivial region as in Figure \ref{fig:EFT_map}.
		The red region has no solution numerically. The label ``Sol:\{a,b\}'' denotes the number of solutions in permutation invariant and cyclic invariant amplitudes respectively.
	}
	\label{fig:num_checks_4pt}
\end{figure}

As was discussed earlier the 4pt amplitudes are special due to 4pt kinematics. All kinematical invariants vanish if we set one of the momenta to zero. Therefore, for $(\partial^m\phi^4)$ we have $\rho=\frac{m-2}{2}$ and $\sigma=\frac{m}{2}$ which implies $\rho=\sigma-1$. But we still have the inequality $\rho<3$ and therefore, the only allowed cases are $(\rho,\sigma)=(0,1),(1,2),(2,3)$. We can now directly explore all these cases with numerical methods and determine how many solutions are in each point of $(\rho,\sigma)$ space. In order to check the existence of such theories we have to perform the test for 6pt amplitudes. The ansatz now contains the factorization terms with 4pt vertices as well as 6pt contact term from the Lagrangian,

\begin{equation}
{\cal L}_\rho = (\partial^{2\rho+2}\phi^4) + (\partial^{4\rho+2}\phi^6).
\end{equation}
We perform the check in $d=3,4,5$ as these are the only interesting cases. The results are summarized in Figure~\ref{fig:num_checks_4pt}. The first solution for $(\rho,\sigma)=(0,1)$ is with cyclic symmetry,
\begin{align}
A_6^{(0,1)} &= \frac{(s_{12}+s_{23})(s_{45}+s_{56})}{s_{123}} + \frac{(s_{23}+s_{34})(s_{56}+s_{61})}{s_{234}} + \frac{(s_{34}+s_{45})(s_{61}+s_{12})}{s_{345}}\nonumber\\
&\hspace{1cm} - (s_{12}+s_{23}+s_{34} + s_{45} + s_{56} + s_{61})\,,
\end{align}
which is the 6pt amplitude in the $SU(N)$ non-linear sigma model in any $d$. The solution for $(\rho,\sigma)=(1,2)$ is with permutational symmetry,

\begin{equation}
A_6^{(1,2)}= \frac{(s_{12}s_{23}+s_{13}s_{23}+s_{12}s_{13})(s_{45}s_{46}+s_{46}s_{56}+s_{45}s_{56})}{s_{123}} - s_{12}s_{34}s_{56} + {\rm permutations}\,,
\end{equation}
which is the 6pt amplitude in the Dirac-Born-Infeld theory in any $d$. The last solution is a 4pt Galileon for $(\rho,\sigma)=(2,2)$ which exists for both single and multiple scalar cases for $d>2$. In the single scalar case there is an extra $\sigma=3$ behavior giving us the special Galileon with $(\rho,\sigma)=(2,3)$ while for the flavor-ordered case this enhanced soft limit is not present.

\subsection{High Valency}

The set of all possible values of $\rho$ is $\rho = \frac{m-2}{v-2}$ where $m$ is the number of derivatives in the interaction and with constraint 
\begin{equation}
3>\frac{m-2}{v-2} \geq \frac{\sigma(v-1)-2}{(v-2)}
\end{equation}
and also $\sigma>\rho$ for $\sigma=1$ and $\sigma\geq \rho$ for $\sigma>1$. These inequalities can be easily solved and we can find all integers $p$ which satisfy them which would enumerate all possible solutions. For $\sigma=1$ it becomes 

\begin{equation}
v> m \geq v-1\,,
\end{equation}
which has the only solution if $m=v-1$. Therefore, the only possible allowed case is $(\rho,\sigma) = (\frac{v-3}{v-2},1)$. 
As this has $\rho<1$ there can not be any permutational invariant amplitude with $\sigma=1$ behavior -- for single scalar the theory must be derivatively coupled. However, we can have cyclically invariant $v$-pt amplitude,
\begin{equation}
A_v^{(\frac{v-3}{v-2},1)} = \epsilon_{\alpha_1\alpha_2\dots \alpha_{v-1}} p_1^{\alpha_1} p_2^{\alpha_2} \dots p_{v-1}^{\alpha_{v-1}}.
\end{equation}
This corresponds to the WZW term which exists only in $d=v-1$ dimensions.
Of course, this is the only possible term if the number of derivatives $m=v-1$ is odd.
For general $v$, we can not prove that the WZW term is the only solution for $v>6$, but all theories have to sit at the point $(\rho,\sigma)=(\frac{v-3}{v-2},1)$.

For $\sigma=2$ the inequality becomes $2(v-2)\geq m\geq 2(v-2)$ which forces $m=2(v-2)$ and $\rho=2$. So the only allowed case is $(\rho,\sigma)=(2,2)$. We know that this is exactly the powercounting of the $v$pt Galileon, in $d=v-1$ dimensions it is
\begin{equation}
A_v^{(2,2)} = (\epsilon_{\alpha_1\alpha_2\dots \alpha_{v-1}} p_1^{\alpha_1} p_2^{\alpha_2} \dots p_{v-1}^{\alpha_{v-1}})^2
\end{equation}
but there is a general form analogous to (\ref{gal5}) in any $d>v-2$. Note that this solution exists for both cyclic and permutational cases. What we can not prove there are no other solutions than Galileon for $v>6$ but they all have to sit at the point $(\rho,\sigma)=(2,2)$.

\subsection*{Exclusion Summary}

To summarize, by direct evaluation we found all possible amplitudes with enhanced soft limit for $v=4,5,6$ which gives all interesting theories for $d=3,4,5$. We found that for $v=4$ these theories are NLSM, DBI, Galileon and WZW theory. For $v=5,6$ we have only Galileon and WZW. Both of these theories exist for $v>6$, and in fact they both populate the only allowed points in $(\rho,\sigma)$ plane. 

As a result, for $v=4,5,6$ we enumerated all such theories and there can not be any new ones. For $v>6$ which is relevant only for $d>5$, there is a possibility new theories can appear but they have to sit in the same $(\rho,\sigma,v,d)$ spots degenerate with WZWs and Galileons.

\section{More Directions}
\label{sec:more}

In this section we discuss several directions not included in the classification above. In particular, we first make some comments about the theories of multiple scalars that cannot be flavor-ordered. We solve this problem for the two flavor case and make some comments about three flavors. The landscape of theories for any number of flavors is still unknown.

We also explore other kinematical limits than just soft limit. In particular, we discuss double soft limit when two momenta go to zero simultaneously, and the collinear limit when two of the momenta become proportional.

\subsection{Multiple Scalars}
\label{sec:multiplescalars}

This analysis exactly mirrors the strategy of \cite{Cheung:2014dqa}, which constructed all single scalar effective theories consistent with factorization and a prescribed value of $(\rho,\sigma)$.  This procedure uniquely lands on well-known theories such as DBI and the Galileon, but also suggested the evidence for a new effective theory known as the so-called special Galileon, whose enhanced shift symmetry is now fully understood \cite{Hinterbichler:2015pqa}.

Here, we apply the same procedure but allow for multiple species.  As this constructive procedure is open ended, we restrict to the simplest case of $N=2$ flavors throughout.  We save $N=3$ and higher to future work.

We start at 4pt, demanding that a general theory of the scalars $\phi_1$ and $\phi_2$ has an enhanced soft limit.  However, we can see that this is automatic, by the following argument.  At a fixed value of the power counting parameter $\rho$, the 4pt amplitude $A_4$ should contain $2(\rho+1)$ powers of momenta, so it is some polynomial in $s,t,u$ with that degree.  As we can always go to a basis that manifests a particular soft limit, {\it e.g.}~the soft limit for leg 1 with $s=p_1 p_2$, $t=p_1 p_4$, $u=p_1 p_3$, then we have that
\bea
A_4 \overset{p\rightarrow 0}{\sim} p^{\rho+1},
\eea
which means that $\sigma = \rho+1$ generically, which corresponds to an enhanced soft limit at 4pt.

To move beyond the 4pt amplitude we must explicitly enumerate the vertices.  First, it is easily seen that any cubic scalar interactions with derivatives can be eliminated via equations of motion, so for example the interaction $\lambda^{(3)}_{ijk} \partial_\mu \phi_i \partial^\mu \phi_j \phi_k$ can be removed by a field redefinition of the form 
\bea
\phi_i \rightarrow \phi_i + \lambda^{(3)}_{ijk} \phi_j \phi_k.
\eea
Thus we can assume the absence of a 3pt vertex.  With interactions that start at the 4pt vertex, the first amplitude of interest is a 6pt, which can receive contribution from the 4pt and 6pt vertex.  

For the two derivative case, $\rho=0$, the general action for $N=2$ flavors is
\bea
{\cal L}_{\rho=0} &=& \frac{1}{2} \partial_\mu \phi_i \partial^\mu \phi_j(\delta_{ij} + \lambda_{ijkl}^{(4)} \phi_k \phi_l + \lambda_{ijklmn}^{(6)} \phi_k \phi_l \phi_m \phi_k  +\ldots),
\eea
without loss of generality.  For $\rho=1$, the general action is
\bea
{\cal L}_{\rho=1} &=& \frac{1}{2} \partial_\mu \phi_i \partial^\mu \phi_j(\delta_{ij} + \lambda_{ijkl}^{(4)} \partial_\nu \phi_k  \partial^\nu \phi_l + \lambda_{ijklmn}^{(6)} \partial_\mu \phi_k \partial^\nu \phi_l \partial_\rho \phi_m \partial^\rho \phi_k  +\ldots)
\eea
and there is a straightforward generalization to $\rho=2$. 

To construct the theory we then computed the 6pt scattering amplitude and demanded $\sigma=1,2,3$ soft limits for the $\rho=0,1,2$ cases.    For $\rho=0$ we find a single solution which corresponds to the $SO(3)/SO(2)$ NLSM, where the $N=2$ flavors correspond to the two massless NGBs.  For $\rho=1$, we find two solutions.  The first solution is simply two copies of the DBI theory for a 4D brane moving in 5D.  The second solution is the DBI theory describing a 4D brane moving in 6D. Finally, for $\rho=2$, the only possible theory in 4D corresponds to the single scalar special galileon.   In these cases the multi-flavor EFTs have the property that they can be rewritten as a sum of independent one-flavor Lagrangians after an orthogonal transformation. As a result, the Feynman rules for vertices are blind to the actual flavor combination of the legs.

\subsection{Double Soft Limits}

To begin, we consider the simultaneous soft limit of two particles, $p_j,p_k\rightarrow 0$.   In the context of the NLSM, this limit is sensitive to the structure of the coset space~\cite{ArkaniHamed:2008gz,Kampf:2013vha,Du:2015esa,Low:2015ogb}, and has been applied in the context of the scattering equations~\cite{Cachazo:2015ksa}. More recently, this kinematic regime was  studied for gauge theory and gravity \cite{Klose:2015xoa,DiVecchia:2015bfa}.

Here we consider the double soft limit for a general scalar EFT.  In this case the distinction between theories with trivial versus non-trivial behavior is different from that of the single soft limit since poles in the denominator can blow up. If $p_1,p_2\rightarrow0$ then all poles $s_{12a}\rightarrow0$ where $a=3,4,\dots,n$. For this reason factorization terms typically are singular, and will not have a smooth double soft limit.

For concreteness, let us consider two momenta, $p_2$, $p_3$, to be sent to zero,
\begin{equation}
p_2(t) = tp_2,\qquad p_3(t) = \alpha t p_3.
\end{equation}
We also shift all other momenta in order to satisfy momentum conservation. The shifted amplitude is then inspected based on the degree of vanishing as $t\rightarrow 0$,
\begin{equation}
A_n(t) = {\cal O}(t^\sigma).
\end{equation}
It is simple to see that 5pt amplitudes are not interesting in this limit since $t\rightarrow0$ yields an on-shell 3pt amplitude which is identically zero by our earlier kinematic arguments. Therefore, the first non-trivial case is the 6pt amplitude, which we now consider in detail.  Furthermore, it is sufficient to fix to $d=5$ for 6pt kinematics. No new solutions can exist for $d>5$ but some of them can disappear when going to $d=4$.   For interesting cases in $d=5$ we check if they are present in $d=4$. While we do not have the similar exclusion bounds as in the single soft limit case---presumably they do exist as well as double soft recursion relations---we can still fix $n=6$ and increase the number of derivatives. 

The first question is what is the meaning of ``non-trivial" from the point of view of the double soft limit. Here it matters critically if we have $v=4$ or $v=6$. For $v=6$ we have only a contact term and therefore $\sigma\geq2\rho$ to get non-trivial soft limit behavior. If we have $v=4$ then there are propagators in factorization terms which blow up for $p_2,p_3\rightarrow0$ and therefore, the behavior is not just naive square of the single scalar soft limit. In particular, we get $\sigma\geq2\rho-1$. In Table~\ref{tab:doublesoft} we summarize the number of solutions for $v=4$ and $v=6$. Note that $v=4$ exist only for integer $\rho$. For $\rho=1$ we have the straight inequalities for a non-trivial bound. 

\begin{table}[tbh]
\begin{center}
	\begin{tabular}{|c|c|c|c|c|c|}
		\hline
		& $\rho=\frac12$& $\rho=1$ & $\rho=\frac32$ & $\rho=2$ & $\rho=\frac52$ \\ \hline 
		$\sigma=1$ & {\color{blue}0} & & & & \\ \hline
		$\sigma=2$ & & $\color{red} {0}$ & &  & \\ \hline
		$\sigma=3$ & &  {\color{blue} 0}&{\color{blue} 0} & $\color{red} {1}$ &  \\ \hline
		$\sigma=4$ & & & & {\color{blue} 1}  &   \\ \hline   
		$\sigma=5$ & & & &  & {\color{blue} 0}  \\ \hline   
	\end{tabular}
\end{center}
\caption{Number of solutions for double soft limit. We denote $\color{red} {n}$ the number of solution for $v=4$ and ${\color{blue} n}$ the number of solutions for $v=6$.}
\label{tab:doublesoft}
\end{table}

We see that there are two interesting cases are for $\rho=2$, one for $v=4$ and one for $v=6$. We can easily identify both of them with Galileons. For $v=4$ it is the 4pt Galileon (which also exists for $d=4$) while for $v=6$ it is 6pt Galileon which is absent in $d=4$ and lower. This can be easily shown from the representation of the Galileon vertex as a Gram determinant. As was shown before each Gram determinant (for any number of points) scales like ${\cal O}(t^2)$ in the soft limit. For $v=6$ we can obtain ${\cal O}(t^4)$ in the double soft limit. For $v=4$ this is reduced by one power due to the propagators (when $p_2$ and $p_3$ are on the same side) which also scales like ${\cal O}(t)$, and in the end we get ${\cal O}(t^3)$. Note that the ${\cal O}(t^3)$ behavior of the special Galileon in the single soft limit is not propagated into the double soft limit case, and only the ${\cal O}(t^2)$ behavior is relevant. Here we performed the checks only for $\rho<3$ but in principle, we should consider higher $\rho$ or prove the same bound as in the single scalar field case.

\subsection{Collinear Limits}

The other natural limit to consider is the collinear limit where two of the momenta become proportional. 
This was recently studied from scattering equations~\cite{Nandan:2016ohb}.
We study it again in the context of single scalar EFT so we can choose $p_3=\alpha p_2$ (for some parameter $\alpha$) without a loss of generality. Unlike the single soft limit and double soft limit cases there are no theoretical expectations how the amplitude should behave. In the Yang-Mills theory and gravity collinear limits are well understood and provide a pole and phase factor, respectively. In our case the situation is different as there are no 3pt vertices and the collinear limit never diverges. Therefore, we can pose the question in a similar way as in the soft limit case: when does the amplitude vanish at a given rate $\sigma$?

To be more specific, we have to introduce a small parameter $t$ which will control the distance from the collinear region. We shift momentum $p_3\rightarrow p_3(t)$ where
\begin{equation}
p_3(t) = \alpha(1-t)p_2 - \alpha t(1-t)\frac{s_{23}}{\alpha(1-t)s_{12} + ts_{13}}p_1 + t p_3,
\label{colshift}
\end{equation}
where $s_{ab}$ are the invariants of unshifted momenta. In order to preserve the momentum conservation we have to shift also other momenta $p_4,\dots,p_n$ but in a way which is regular for any value of $t$. The shift in \Eq{colshift} is more complicated in order to preserve the on-shell condition $p_3(t)^2=0$ and also control the way how we approach the collinear region. Note that for $t=1$ we recover the original configuration, $p_3(t)=p_3$ and also other momenta become unshifted, while for $t=0$ we get $p_3=\alpha p_2$. Then the question is what is the rate at which the shifted amplitude $A_n(t)$ vanishes,
\begin{equation}
A_n(t) = {\cal O}(t^\sigma).
\end{equation}

Unlike in the soft limit case there is no statement symmetry $\rightarrow$ collinear limit. Therefore, we have to rely just on the kinematical check. The only kinematical invariant which vanishes in this limit is $s_{23}$. Naively, in order to get the vanishing collinear limit in any pair of momenta each Feynman diagram would have to contain the product of all invariants $s_{ij}$ which pushes the derivative degree very high. We also do not have any argument about the leading valency of the Lagrangian.

We did the checks for 5pt amplitudes up to 18 derivatives, 6pt amplitudes up to 14 derivatives and 7pt amplitudes up to 10 derivatives, with no interesting results (no vanishing collinear limits) except one class of theories which are Galileons.

\subsubsection*{Galileons from collinear limits}

For the Lagrangians of the type $(\partial^8\phi^5)$ there is one solution for the collinear limit vanishing for $d\geq 4$, and for $(\partial^{10}\phi^6)$ there is also one solution for $d\geq5$. The solutions can be identified with the 5pt and 6pt Galileons which are then unique solutions to the problem of vanishing collinear limit. Moreover, the amplitudes in both cases vanish as $A(t)\sim {\cal O}(t^2)$. This can be understood from the definition of the Galileon vertex. The Gram determinant for $n=5$ in $d=4$ behaves by definition as 
\begin{equation}
{\rm Gram}_{d=4,n=5}\left[p_1,p_2,p_3(t),p_4(t),p_5(t)\right] \equiv (\epsilon_{\mu\nu\alpha\beta} p_{1\mu}p_{2\nu}p_{3\alpha}(t)p_{4\beta}(t))^2 = {\cal O}(t^2) \label{GramCol}
\end{equation}
and similarly for $n=6$ and $d=5$. In higher dimensions some of indices are contracted together from both $\epsilon$ tensors but the scaling property is still valid. However, the collinear vanishing is the property of the contact term only, not the amplitude for higher $n$. The factorization terms spoil this property as they lack do not vanish in the collinear limit when both legs are on the opposite sides of the channel. In principle, there could be a cancellation between different Feynman diagrams, but this does not happen as the numerical checks show.  We can also see it in the $(\partial^{10}\phi^6)$ case where there is no solution for the 6pt amplitude coming from the 4pt Galileon $(\partial^6\phi^4)$. 

But still it is interesting to note that the collinear limit can be used to define the Galileons as unique theories based on the behavior in the collinear limit. It would be interesting to explore the kinematical space more exhaustively and also do it for multiple scalars.




\section{Outlook}
\label{sec:conclusion}

In this paper we have mapped out the theory space of Lorentz invariant and local scalar effective field theories by studying the soft behavior of scattering amplitudes. The bulk of our discussion has focused on theories of a single scalar or multiple scalars which allow for flavor-ordering. We have derived bounds on the power counting and soft behavior of all possible consistent theories with enhanced soft limit and classified completely all the non-trivial cases in $d<6$. Our final catalog of EFTs include NLSM, DBI, Galileon, and WZW term theory. 
A main takeaway of this paper is that these theories are truly unique.  We also commented on the theories with generic multiple scalars and different kinematical limits. 

Remarkably, the exceptional theories discussed here coincide precisely with the EFTs constructed from the CHY representation~\cite{Cachazo:2014xea} and which satisfy BCJ duality~\cite{Chen:2013fya}.
Moreover, there is evidence of new theories which are extensions of these exceptional theories~\cite{Cachazo:2016njl,Carrasco:2016ldy}, suggesting a rich interplay between soft limits, BCJ duality, and CHY representation.
Classifying theories based on various aspects can illuminate the relations among them. 
Insights into the soft structure of the S-matrix have also arisen in the program of asymptotic symmetries~\cite{Strominger:2013lka,He:2014cra,He:2015zea,Kapec:2015ena,Strominger:2015bla,Bondi:1962px,Sachs:1962wk,Barnich:2009se,Strominger:2013jfa,He:2014laa,Kapec:2014opa,Lysov:2014csa,Dumitrescu:2015fej,Weinberg:1965nx,Low:1958sn,Burnett:1967km,White:2011yy,Cachazo:2014fwa,Cheung:2016iub}.

There are many other directions viable for constructing theories from the properties of scattering amplitudes. The most natural directions is to consider other particle content (higher spins), other kinematical regimes (like double soft limit or collinear limit briefly mentioned in the paper), loop-level correction~\cite{Goon:2016ihr}, or curved backgrounds.
More ambitiously, one might also consider non-relativistic theories~\cite{Griffin:2014bta}, where amplitudes satisfy less symmetry, but must nevertheless exhibit locality and factorization.   A priori, one would expect a far greater diversity in non-relativistic EFTs, so there is also the possibility that new theories might yet lay undiscovered.

This is the first step in the program of extending the developments in the study of scattering amplitudes in gauge theory and gravity to other quantum field theories, and EFTs are the furthest possible cousins. 
The recent progress on recursion relations and CHY representation in these theories show that there should be a completely new formulation for scattering in general QFTs.

\acknowledgments
We thank Nima Arkani-Hamed and Enrico Herrmann for discussions. We also thank Tom Melia for pointing out a typo in the earlier version. This work is supported in part by Czech Government project no.~LH14035 and GACR 15-18080S. C.C.~and C.-H.S.~are supported by a Sloan Research Fellowship and a DOE Early Career Award under grant no.~DE-SC0010255. C.-H.S.~thanks Nordita and Niels Bohr International Academy and Discovery Center for hospitality.

\appendix

\section{Proof of the Soft Theorem}
\label{app:proofsoft}

In this Appendix we give detailed proof of the soft theorem mentioned in the
Section 3. While the bulk of this paper focuses on tree-level scattering
amplitudes, we present here a non-perturbative proof which to our knowledge
does not exists in the literature. For simplicity we restrict ourselves to
theory with single NGB, while the generalization to multiple flavors is
straightforward.


\subsection*{Review of the Adler Zero}


For our analysis it will be helpful to briefly review the derivation of the
Adler zero for the amplitudes of NGBs (see \textit{e.g.}~the textbook~\cite%
{Weinberg:1996kr} and references therein). To begin, consider a theory of a
single NGB corresponding to the spontaneous breaking of a one-parameter
continuous symmetry. In most cases such a symmetry acts non-linearly on the
NGB field according to 
\begin{equation}
\phi \left( x\right) \rightarrow \phi \left( x\right) +a ,  \label{shift}
\end{equation}%
which has an associated Noether current $J^\mu(x)$. The NGB couples to the
current with a strength parameterized by the decay constant, $F$, so 
\begin{equation}
\left\langle 0|J^{\mu }(x)|\phi (\mathbf{p})\right\rangle = i p^{\mu
}Fe^{-ip\cdot x}.
\end{equation}%
The matrix elements of the current $J^{\mu }(x)$ has a pole as $%
p^{2}\rightarrow 0$ whose residue is related to the amplitude for the NGB
emission,\footnote{%
Here and in what follows we tacitly assume that all the momentum
conservation $\delta -$functions are removed form the matrix elements. I.e. $%
R^{\mu }$ does not contain momentum conservation $\delta -$functions.} 
\begin{eqnarray}
\left\langle \alpha ,\mathrm{out}|J^{\mu }(0)|\beta ,\mathrm{in}%
\right\rangle &=&\frac{i}{p^{2}}\left\langle 0|J^{\mu }(0)|\phi (\mathbf{p}%
)\right\rangle \left\langle \alpha +\phi (\mathbf{p}),\mathrm{out}|\beta ,%
\mathrm{in}\right\rangle +R^{\mu }(p)  \nonumber \\
&=&-\frac{p^{\mu }}{p^{2}}F\left\langle \alpha +\phi (\mathbf{p}),\mathrm{out%
}|\beta ,\mathrm{in}\right\rangle +R^{\mu }(p)  \label{J_pole}
\end{eqnarray}%
where $p^\mu=P^\mu_{\beta,{\mathrm{in}}}-P^\mu_{\alpha, {\mathrm{out}}}$ is
the difference in the in and out momenta, and $R^{\mu }(p)$ denotes a
remainder function which is regular as $p^{2}\rightarrow 0$. Due to
conservation of $J^{\mu }$ we can dot Eq.~(\ref{J_pole}) into $p^\mu$ to
obtain the equation 
\begin{equation}
\left\langle \alpha +\phi (\mathbf{p}),\mathrm{out}|\beta ,\mathrm{in}%
\right\rangle =\frac{1}{F}p_{\mu }R^{\mu }(p),  \label{amplitude}
\end{equation}%
so $p_{\mu }R^{\mu }(p)/F$ can be thought of as an off-shell extension of
the amplitude. The behavior of the amplitude in the soft NGB limit $%
p\rightarrow 0$ can be therefore inferred from the properties of the
remainder function $R^{\mu }(p)$. Provided the theory does not have a cubic
vertex, then $R^{\mu }(p)$ is regular for $p\rightarrow 0$, which implies
that 
\begin{equation}
\lim_{p\rightarrow 0}\left\langle \alpha +\phi (\mathbf{p}),\mathrm{out}%
|\beta ,\mathrm{in}\right\rangle =\frac{1}{F}\lim_{p\rightarrow 0}p_{\mu
}R^{\mu }(p)=0 .  \label{adler_zero}
\end{equation}%
This condition is precisely the Adler zero for NGB soft emission.


\subsection*{Classical Current Relations}


It is straightforward to extend our results to the case of a generalized
shift symmetry, 
\begin{equation}
\phi \rightarrow \phi +\delta _{\theta }\phi \left( x\right)
\label{symmetry}
\end{equation}
where the variation takes the form 
\begin{equation}
\delta _{\theta }\phi \left( x\right) =\theta _{j}\alpha _{A}^{j}(x) O^{A} 
\left[ \phi \right] \left( x\right) .  \label{transformation}
\end{equation}%
Here $\theta _{j}$ are infinitesimal parameters, $\alpha _{A}^{j}\left(
x\right) $ are fixed polynomial functions, $O^{A}\left[ \phi \right] \left(
x\right) $ are local but generally composite operators constructed from $%
\phi \left( x\right) $ and its derivatives.

Classically, we can consider the local shift transformation, $\phi
(x)\rightarrow \phi (x)+a(x)$, with a shift parameter with special value of $%
a\left( x\right) =\widehat{a}\left( x\right) $, namely with%
\[
\widehat{a}\left( x\right) =\theta _{j}\left( x\right) \alpha _{A}^{j}\left(
x\right) O^{A}\left[ \phi \right] \left( x\right) ,
\]%
which coincides with the localized version of the transformation Eq.~(\ref%
{transformation}) with parameters $\theta _{j}\rightarrow \theta _{j}\left(
x\right) $. This induces a relation between the Noether current of the shift
symmetry $J^{\mu }\left( x\right) $ and the Noether current $J^{\left(
j\right) \mu }\left( x\right) $ corresponding to the transformation Eq.~(\ref%
{transformation}) see\cite{Brauner:2014aha} for general discussion and
further details)%
\begin{equation}
\int d^{d}x\;\partial \widehat{a}\cdot J=\int d^{d}x\;\partial \theta
_{j}\cdot J^{\left( j\right) }\left( x\right) .
\end{equation}%
Explicitly, we obtain%
\begin{equation}
\int d^{d}x\left[ \partial \theta _{j}\alpha _{A}^{j}O^{A}\left[ \phi \right]
+\theta _{j}\partial \alpha _{A}^{j}O^{A}\left[ \phi \right] +\theta
_{j}\alpha _{A}^{j}\partial O^{A}\left[ \phi \right] \right] \cdot J=\int
d^{d}x\;\partial \theta _{j}\cdot J^{\left( j\right) }\left( x\right) .
\label{nc}
\end{equation}%
Invariance of the action with respect to the global form of the
transformation Eq.~(\ref{transformation}) means that for constant $\theta
_{j}$, the integrand on the left-hand side of the previous equation is a
total derivative 
\[
\left( \partial \alpha _{A}^{j}O^{A}\left[ \phi \right] +\alpha
_{A}^{j}\partial O^{A}\left[ \phi \right] \right) \cdot J=\partial _{\alpha
}\left( \beta _{I}^{\alpha j}\mathcal{O}^{I}\left[ \phi \right] \right) ,
\]%
where as above $\beta _{I}^{j}$ are known functions and $\mathcal{O}^{I}$are
local composite operators. Inserting the latter into Eq.~(\ref{nc}) we get%
\[
\int d^{d}x\;\partial \theta _{j}\cdot J^{\left( j\right) }\left( x\right)
=\int d^{d}x\left[ J\cdot \partial \theta _{j}\alpha _{A}^{j}O^{A}\left[
\phi \right] +\theta _{j}\partial _{\alpha }\left( \beta _{I}^{\alpha j}%
\mathcal{O}^{I}\left[ \phi \right] \right) \right] 
\]%
and thus%
\[
J^{\left( j\right) \mu }=\alpha _{A}^{j}O^{A}\left[ \phi \right] J^{\mu
}-\beta _{I}^{\mu j}\mathcal{O}^{I}\left[ \phi \right] .
\]%
To summarize, we get two algebraic off-shell identities%
\begin{eqnarray}
\left( \partial \alpha _{A}^{j}O^{A}\left[ \phi \right] +\alpha
_{A}^{j}\partial O^{A}\left[ \phi \right] \right) \cdot J &=&\partial \cdot
\beta _{I}^{j}\mathcal{O}^{I}\left[ \phi \right] +\beta _{I}^{j}\cdot
\partial \mathcal{O}^{I}\left[ \phi \right]   \nonumber \\
J^{\left( j\right) } &=&\alpha _{A}^{j}O^{A}\left[ \phi \right] J-\beta
_{I}^{j}\mathcal{O}^{I}\left[ \phi \right] ,  \label{relation}
\end{eqnarray}%
which reveals the underlying dependence between the currents: conservation
of $J^{\left( j\right) }$ is a consequence of conservation of $J$.

Let us now apply these relation to the case when $O^{1}\left[ \phi \right]
=1 $, i.e. when we can rewrite Eq.~(\ref{transformation}) in the form 
\begin{equation}
\delta _{\theta }\phi \left( x\right) =\theta _{j}\left[ \alpha ^{j}\left(
x\right) +\alpha _{B}^{j}\left( x\right) O^{B}\left[ \phi \right] \left(
x\right) \right]  \label{transofrmation_1}
\end{equation}%
Such a transformation can be understood as a generalization of the simple
shift symmetry Eq.~(\ref{shift}) or more generally of the polynomial shift
symmetry discussed in \cite{Griffin:2014bta} and \cite{Hinterbichler:2014cwa}%
. Note again that $\alpha ^{j}\left( x\right) $ and $\alpha _{B}^{j}\left(
x\right) $ are polynomials. Then the first of the relations Eq.~(\ref%
{relation}) reads 
\begin{equation}
\partial \alpha ^{j}\cdot J=-\partial \cdot \left( \alpha _{B}^{j}O^{B}\left[
\phi \right] J-\beta _{I}^{j}\mathcal{O}^{I}\left[ \phi \right] \right)
+\alpha _{B}^{j}O^{B}\left[ \phi \right] \partial \cdot J
\end{equation}
From now we will assume just this special form of the relation between
currents.

\subsection*{Quantum Current Relations}

Another important assumption is that above mentioned relations survive
quantization, so for the renormalized quantum operators we have the current
conservation equation, 
\[
\partial \cdot \left\langle \alpha ,\mathrm{out}|J^{\left( j\right)
}(x)|\beta ,\mathrm{in}\right\rangle =\partial \cdot \left\langle \alpha ,%
\mathrm{out}|J(x)|\beta ,\mathrm{in}\right\rangle =0, 
\]%
as well as the relation 
\begin{eqnarray}
\partial \alpha ^{j}(x)\cdot \left\langle \alpha ,\mathrm{out}|J(x)|\beta ,%
\mathrm{in}\right\rangle &=&-\partial \cdot \left\langle \alpha ,\mathrm{out}%
|\alpha _{B}^{j}\left( x\right) O^{B}\left[ \phi \right] \left( x\right)
J\left( x\right) -\beta _{I}^{j}\left( x\right) \mathcal{O}^{I}\left[ \phi %
\right] \left( x\right) |\beta ,\mathrm{in}\right\rangle  \nonumber \\
&&+\alpha _{B}^{j}\left( x\right) \left\langle \alpha ,\mathrm{out}|O^{B}%
\left[ \phi \right] \left( x\right) \partial \cdot J\left( x\right) |\beta ,%
\mathrm{in}\right\rangle .  \label{crutial_identity}
\end{eqnarray}
Evaluated between on-shell in and out states, we obtain%
\begin{equation}
\left\langle \alpha ,\mathrm{out}|O^{B}\left[ \phi \right] \left( x\right)
\partial \cdot J\left( x\right) |\beta ,\mathrm{in}\right\rangle =0
\label{current_conservation}
\end{equation}%
as a consequence of the Ward identities for the current $J$. Therefore, 
\[
\partial \alpha ^{j}(x)\cdot \left\langle \alpha ,\mathrm{out}|J(x)|\beta ,%
\mathrm{in}\right\rangle =\partial \cdot \left\langle \alpha ,\mathrm{out}%
|\gamma _{C}^{j}\left( x\right) \mathfrak{O}^{C}\left[ \phi \right] \left(
x\right) |\beta ,\mathrm{in}\right\rangle, 
\]%
where we denoted collectively all the $c-$number functions $\alpha
_{B}^{j}\left( x\right) $ and $\beta _{I}^{j}\left( x\right) $ as $\gamma
_{C}^{j}\left( x\right) $ and the local operators $O^{B}\left[ \phi \right]
\left( x\right) J\left( x\right) $ and $\mathcal{O}^{I}\left[ \phi \right]
\left( x\right) $ as $\mathfrak{O}^{C}\left[ \phi \right] \left( x\right) $.
We then obtain 
\begin{equation}
e^{-ip\cdot x}\partial \alpha ^{j}\left( x\right) \cdot \left\langle \alpha ,%
\mathrm{out}|J(0)|\beta ,\mathrm{in}\right\rangle =\partial \cdot \left[
\gamma _{C}^{j}\left( x\right) e^{-ip\cdot x}\right] \left\langle \alpha ,%
\mathrm{out}|\mathfrak{O}^{C}\left[ \phi \right] \left( 0\right) |\beta ,%
\mathrm{in}\right\rangle  \label{JO_quantum}
\end{equation}%
with $p=P(\beta _{\mathrm{in}})-P(\alpha _{\mathrm{out}})$ for any in and
out states. For special choice $\langle \alpha ,\mathrm{out}|=\langle 0|$
and $|\beta ,\mathrm{in}\rangle =|\phi ^{i}(\mathbf{p})\rangle $ we get 
\begin{equation}
\partial \alpha ^{j}\left( x\right) \cdot \left\langle 0|J(0)|\phi (\mathbf{p%
})\right\rangle =\partial \cdot \left[ \gamma _{C}^{j}\left( x\right)
e^{-ip\cdot x}\right] \left\langle 0|\mathfrak{O}^{C}\left[ \phi \right]
\left( 0\right) |\phi (\mathbf{p})\right\rangle .  \label{JO_one_particle}
\end{equation}%
Since the left-hand side of Eq.~(\ref{JO_quantum}) has a NGB pole for $%
p^{2}\rightarrow 0$, this must be reproduced on the right-hand side.
Therefore at least one matrix element $\left\langle \alpha ,\mathrm{out}|%
\mathfrak{O}^{C}\left[ \phi \right] \left( 0\right) |\beta ,\mathrm{in}%
\right\rangle $ develops a pole. In general we can write%
\begin{equation}
\left\langle \alpha ,\mathrm{out}|\mathfrak{O}^{C}\left[ \phi \right] \left(
0\right) |\beta ,\mathrm{in}\right\rangle =\frac{i}{p^{2}}\left\langle 0|%
\mathfrak{O}^{C}\left[ \phi \right] \left( 0\right) |\phi (\mathbf{p}%
)\right\rangle \left\langle \alpha +\phi (\mathbf{p}),\mathrm{out}|\beta ,%
\mathrm{in}\right\rangle +R^{C}(p)  \label{O_pole}
\end{equation}%
where $R^{C}(p)$ is a remnant regular for $p^{2}\rightarrow 0$ and therefore
at least one matrix element $\left\langle 0|\mathfrak{O}^{C}\left[ \phi %
\right] \left( 0\right) |\phi ^{i}(\mathbf{p})\right\rangle $ must be
nonzero.

Inserting Eq.~(\ref{J_pole}) and Eq.~(\ref{O_pole}) into Eq.~(\ref%
{JO_quantum}), together with Eq.~(\ref{JO_one_particle}), we obtain the
following relation between the remainder functions 
\begin{equation}
e^{-ip\cdot x}\partial \alpha ^{j}\left( x\right) \cdot R(p)=\partial \cdot 
\left[ \gamma _{C}^{j}\left( x\right) e^{-ip\cdot x}\right] R^{C}(p).
\end{equation}%
In what follows we will assume that all the remnants are regular also for $%
p\rightarrow 0$, i.e. there are no problems with cubic vertices.

Integrating this over $\mathrm{d}^{d}x$ we get in the sense of distributions%
\begin{equation}
\widetilde{\partial \alpha ^{j}}(p)\cdot R(p)=i\widetilde{\alpha ^{j}}%
(p)p\cdot R(p)=0,  \label{soft_theorem_master_formula}
\end{equation}%
where the tildes denote Fourier transform. Because $p\cdot R(p)$ is related
to the amplitude via Eq.~(\ref{amplitude}), we can infer additional
information on the soft behavior of the amplitude on top of Eq.~(\ref%
{adler_zero}). As we will see in the next subsection, Eq.~(\ref%
{soft_theorem_master_formula}) is the key formula for deriving the soft
theorems for NGBs. Let us note, that it depends only on the $c-$number part
of the general symmetry transformation Eq.~(\ref{transofrmation_1}).
Therefore, theories invariant with respect to the transformation Eq.~(\ref%
{transofrmation_1}) with the same $\alpha ^{j}\left( x\right) $ form
universality classes with the same soft behavior. In the next subsection we
will illustrate application of this formula in more detail.

\subsection*{Derivation of Soft Theorems}

As shown above, the existence of a non-linearly
realized shift symmetry in Eq.~(\ref{shift}) together with the absence of
cubic vertices implies the presence of the Adler zero, \textit{i.e.}~that
the amplitude with one soft emission behaves at least as $\mathcal{O}\left(
p\right) $ for $p\rightarrow 0$.

This result and the case when for $\alpha \left( x\right) =\theta \cdot x$
mentioned in the main text can be easily generalized for the class of
theories invariant with respect to the generalized polynomial shift
symmetries 
\begin{equation}
\delta _{\theta }\phi \left( x\right) =\theta _{\alpha _{1}\ldots \alpha
_{n}}\left[ x^{\alpha _{1}}\ldots x^{\alpha _{n}}+\alpha _{B}^{\alpha
_{1}\ldots \alpha _{n}}\left( x\right) O^{B}\left[ \phi \right] \left(
x\right) \right] ,  \label{generalized_polynomial_shift}
\end{equation}%
which corresponds to $\alpha ^{j}\left( x\right) \rightarrow \alpha ^{\alpha
_{1}\ldots \alpha _{n}}(x)\equiv x^{\alpha _{1}}\ldots x^{\alpha _{n}}$.
Instead of Eq.~(\ref{linear_shift_class}) we get in this case 
\begin{eqnarray}
0 &=&p_{\mu }R^{\mu }(p)\partial ^{\alpha _{1}}\ldots \partial ^{\alpha
_{n}}\delta ^{(4)}(p)  \nonumber \\
&=&\sum_{k=0}^{n}\left( 
\begin{array}{c}
n \\ 
k%
\end{array}%
\right) \left( -1\right) ^{k}\left[ \lim_{p\rightarrow 0}\partial ^{\alpha
_{1}}\ldots \partial ^{\alpha _{k}}p_{\mu }R^{\mu }(p)\right] \partial
^{\alpha _{k+1}}\ldots \partial ^{\alpha _{n}}\delta ^{(4)}(p)
\label{soft_theorem_for_polynomial_shift}
\end{eqnarray}%
and thus for $k=0,\ldots ,n$ 
\begin{equation}
\lim_{p\rightarrow 0}\partial ^{\alpha _{1}}\ldots \partial ^{\alpha
_{k}}p_{\mu }R^{\mu }(p)=0.
\end{equation}%
Using the correspondence in Eq.~(\ref{amplitude}) we conclude that the
amplitude has $\mathcal{O}\left( p^{n+1}\right) $ soft behavior, \textit{i.e.%
}~an Adler zero of the $(n+1)$th order.

It is also straightforward to generalize the above result to the case of
symmetries in Eq.~(\ref{generalized_polynomial_shift}) with traceless tensor 
$\theta _{\alpha _{1}\cdots \alpha _{n}}$. The special Galileon is a member
of this class, and is symmetric with respect to the ``hidden galileon
symmetry''~\cite{Hinterbichler:2015pqa} (see also Appendix~\ref%
{app:sumoftheories}) 
\[
\delta _{s}\phi \left( x\right) =\theta _{\alpha \beta } \alpha^2 x^{\alpha
}x^{\beta }-\theta _{\alpha \beta }\partial ^{\alpha }\phi \left( x\right)
\partial ^{\beta }\phi \left( x\right) 
\]%
where $\theta _{\alpha \beta }=\theta _{\beta \alpha }$ satisfies $\theta
_{\alpha }^{\alpha }=0$. Instead of rewriting the general formula Eq.~(\ref%
{soft_theorem_for_polynomial_shift}) for traceless tensor $\theta _{\alpha
_{1}\ldots \alpha _{n}}$we will illustrate it just on this concrete example.
In this case we have from Eq.~(\ref{transofrmation_1}) 
\[
\alpha ^{j}\left( x\right) \rightarrow \alpha ^{\mu \nu }\left( x\right)
=x^{\mu }x^{\nu }-\frac{1}{d}x^{2}\eta ^{\mu \nu }. 
\]%
Taking the Fourier transform, we obtain 
\begin{eqnarray}
\widetilde{\alpha ^{\mu \nu }}\left( p\right) &=-(2\pi )^{d}\Pi ^{\mu \alpha
\nu \beta }\partial _{\alpha }\partial _{\beta }\delta ^{(d)}(p) \\
\Pi ^{\mu \alpha \nu \beta }&=\eta ^{\mu \alpha }\eta ^{\nu \beta }-\frac{1}{%
d}\eta ^{\mu \nu }\eta ^{\alpha \beta }
\end{eqnarray}%
which with Eq.~(\ref{soft_theorem_master_formula}) implies that 
\begin{eqnarray*}
0 &=&-p_{\sigma }R^{\sigma }(p)\Pi ^{\mu \alpha \nu \beta }\partial _{\alpha
}\partial _{\beta }\delta ^{(d)}(p) \\
&=&-\Pi ^{\mu \alpha \nu \beta }\left\{ \left[ \partial _{\alpha }\partial
_{\beta }\delta ^{(d)}(p)\right] \left[ \lim_{p\rightarrow 0}p_{\sigma
}R^{\sigma }(p)\right] -\left[ \partial _{\alpha }\delta ^{(d)}(p)\right] %
\left[ \lim_{p\rightarrow 0}\partial _{\beta }p_{\sigma }R^{\sigma }(p)%
\right] \right. \\
&&\left. -\left[ \partial _{\beta }\delta ^{(d)}(p)\right] \left[
\lim_{p\rightarrow 0}\partial _{\alpha }p_{\sigma }R^{\sigma }(p)\right]
+\delta ^{(d)}(p)\left[ \lim_{p\rightarrow 0}\partial _{\alpha }\partial
_{\beta }p_{\sigma }R^{\sigma }(p)\right] \right\} .
\end{eqnarray*}%
We have thus soft theorems in the form\footnote{%
This equation also appears in \cite{Hinterbichler:2015pqa}.}%
\begin{equation}
\lim_{p\rightarrow 0}\left( \eta ^{\mu \alpha }\eta ^{\nu \beta }-\frac{1}{d}%
\eta ^{\mu \nu }\eta ^{\alpha \beta }\right) \partial _{\alpha }\partial
_{\beta }\left\langle \alpha +\phi (\mathbf{p}),\mathrm{out}|\beta ,\mathrm{%
in}\right\rangle =0.
\end{equation}%
Taking the soft NGB momentum to be on-shell, we see that the soft limit
vanishes with two powers of momenta, leaving $\mathcal{O}\left( p^{3}\right) 
$ behavior for the amplitude.

To summarize, the soft theorems above hold for an EFT that is invariant with
respect to the generalized polynomial shift symmetry in Eq.~(\ref{symmetry}%
). On the quantum level this means that the relations in Eq.~(\ref%
{crutial_identity}) and Eq.~(\ref{current_conservation}) apply. Note that at
tree-level, the relations Eq.~(\ref{crutial_identity}) and (\ref%
{current_conservation}) are satisfied automatically and therefore the
symmetry (and the absence of the cubic vertices) provides us with a
sufficient condition for enhanced soft limit of the tree-level amplitudes.

\section{Bounds on \texorpdfstring{$\rho$}{rho} from Bonus Relations}
\label{app:proofbonus}
This appendix shows how to obtain rigorous bounds on the power counting parameter $\rho$ in non-trivial theories from bonus relations. We first introduce bonus relations in recursion and then apply them to the spurious pole cancellation.

In normal recursion relations, inputs from all factorization channels are needed.
However, for sufficiently high $\sigma$, it is possible to eliminate certain factorization channels from the recursion relation by introducing factors like $B(z) = P^2(z)/P^2(0)$ directly into the recursion relation
\begin{equation}
\oint \frac{dz}{z}\frac{A_6(z) }{F(z) } B(z).
\label{eq:Cauchy_bonus_raw}
\end{equation}
These terms evaluate to unity at $z=0$, and do not spoil large $z$ behavior, provided the soft behavior is sufficiently enhanced. To isolate the spurious pole cancellation, we choose $B(z)=P^2_{124}(z) P^2_{125}(z) P^2_{126}(z)/P^2_{124} P^2_{125} P^2_{126}$ such that the spurious pole of $a_1-a_2$ only appears in the channel $P^2_{123}(z)=0$.
It relies on the fact that $A_6(z)/F(z)$ vanishes faster then $1/z^6$, 
\begin{equation}
\text{Bonus relation: }\left \lbrace
\begin{array}{lc}
\text{Exceptional theory:} & \rho  \ge 4 \\
\text{Non-trivial theory:} & \rho \ge 5
\end{array}
\right.
\label{eq:bonus_condition}
\end{equation}
which must be satisfied in order to eliminate these factorization channels from the recursion.

We can identify the spurious pole using ``bonus'' recursion relations as the derivation for Eq.~\eqref{eq:leading_spurious_normal}. The only difference is the extra factor of $B(z)$ which is proportional to $f_1^3(z)$ when taking the residue at $z=1/a_2$. Since there is only one single term, we can drop all overall kinematic invariants and the spurious pole becomes
\begin{equation}
\begin{split}
\frac{\lambda_{3,b_{\rm max}}A_L(z)}
{(a_1-a_2)^{b_{\rm max}-2\Delta-3}},
\end{split}
\label{eq:leading_spurious_bonus}
\end{equation}
where the spurious pole power is shifted by 3 from $B(z)$. 
This has to vanish identically when $b_{\rm max}-2\Delta-3$. We discuss the single and multiple scalars in turn.

For single scalar, there is no state sum and $A_L(z)$ can be dropped. As in Eq.~\eqref{eq:bound_b}, we find $2\Delta+3 \ge b_{\rm max} \ge \rho+1-(2\Delta+3)$ which can be satisfied for
\begin{equation}
\left \lbrace
\begin{array}{lc}
\text{Exceptional theory:} & \rho  \le 5 \\
\text{Non-trivial theory:} & \rho \le 9
\end{array}
\right.
\label{eq:bound_b_bonus1}
\end{equation}
These rigorous bounds truncate the range of numerical checks on the spurious pole cancellation in Eq.~\eqref{eq:leading_spurious_normal}.
In the case of stripped amplitudes, we only need to eliminate two factorization channels which is viable for $\rho >0$.
Specifically, choosing $B(z)=P^2_{234}(z)P^2_{612}(z)/P^2_{234}P^2_{345}$ yields spurious pole in $a_1-a_2$ unless $b_{\rm max} \le 1$. This rigorous derivation matches the previous numerical evidence. So we still conclude that the NLSM is the unique exceptional theory with stripped amplitudes.

For multiple scalars, plugging ansatze in Eq.~\eqref{eq:4pt_ansatz_multi} into Eq.~\eqref{eq:leading_spurious_bonus} gives
\begin{equation}
\begin{split}
\frac{\sum_{b'}\lambda^{123}_{b_{\rm max},b'} s_{\hat{4}\hat{5}}^{b'}\, s_{\hat{4}\hat{6}}^{\rho+1-b'}}
{(a_1-a_2)^{b_{\rm max}-2\Delta-3}}
\end{split}
\label{eq:leading_spurious_multi}
\end{equation}
Note that the 4pt kinematics $\hat{p}_{3,4,5,6}$ is generic.
Since the momenta $p_{3,4,5,6}$ are only constrained by 6pt kinematics with $p_{1,2}$, they are sufficient to construct generic 4pt kinematics under the shift.
The two Mandelstam variables $s_{\hat{4}\hat{5}},s_{\hat{4}\hat{6}}$ are therefore independent.
The vanishing of the spurious pole then requires $\lambda^{123}_{b,b'}=0$ unless $b\le 2\Delta+3$ for any $b'$. The bounds are the same as in the single scalar case, Eq.~\eqref{eq:bound_b_bonus1}.

In sum, bonus relations rigorously constrain the upper limits of $\rho$. This is supplementary to the numerical checks of Eq.~\eqref{eq:leading_spurious_normal} which applies to lower $\rho$ then Eq.~\eqref{eq:bonus_condition}.
Combining the two establish the proof of $\rho<3$ for all non-trivial theories, independent of the flavor structure.

\section{Catalog of Scalar Effective Field Theories}
\label{app:sumoftheories}

Here we list known scalar EFTs and their Lagrangians.  These theories typically have generalized shift symmetries, and most have  non-trivial soft behavior in scattering amplitudes.

\subsection*{Non-linear Sigma Model and WZW Term}

The $SU(N)$ non-linear sigma model can be defined by the following Lagrangian

\begin{equation}
{\cal L} = \frac{F^2}{4}{\rm Tr}\,(\partial^\mu U \partial_\mu U^\dagger),\quad\mbox{where}\quad U = \exp\left(\frac{i}{F}\phi\right)
\end{equation}
where $\phi=\phi^aT^a$ is the $(N^2-1)$-plet (octet for $N=3$) of pseudoscalar mesons. The Lagrangian is invariant under the chiral symmetry $U(x)\to V_R U(x) V_L^\dagger$ with unitary matrices $V_{R,L}$. The axial part of this symmetry is realized non-linearly as $\phi\rightarrow \phi + a + \ldots$ where the ellipses stand for  terms that are at least quadratic in field $\phi$ and this implies that the axial symmetry is spontaneously broken. Following the theorem in Sec..., the soft limits of scattering amplitudes vanish, $A = {\cal O}(p)$. This theory for $N=2,3$ is famously used for the description of low energy degrees of freedom of QCD. 

The other theory of this kind involving the same multiple of particles is the following Lagrangian
\begin{equation}\label{eq:wzw4}
{\cal L} = \frac{1}{4} {\rm Tr}(\partial_\mu\phi \partial^\mu \phi) + \lambda \epsilon_{\mu\nu\alpha\beta}\,{\rm Tr}(\phi\,\partial^\mu \phi\,\partial^\nu \phi\,\partial^\alpha\phi\,\partial^\beta \phi)
\end{equation}
It possesses the shift symmetry $\phi\to\phi+a$ and has thus the $O(p)$ behavior. This Lagrangian can be obtained as $\phi\to0$ limit of the famous Wess-Zumino-Witten term
\begin{equation}
S_{WZW} = i\lambda \epsilon^{ABCDE}\int d^5x {\rm Tr}(U^\dagger \partial_A U U^\dagger \partial_B UU^\dagger \partial_C UU^\dagger \partial_D UU^\dagger \partial_E U)\,,
\end{equation}
which corresponds to the chiral anomaly.
Generalization of \eqref{eq:wzw4} beyond $d=4$ is obvious 
\begin{equation}
{\cal L} = \frac{1}{4} {\rm Tr}(\partial_\mu\phi \partial^\mu \phi) + \lambda \epsilon_{\mu_1 \ldots \mu_d}\,{\rm Tr}(\phi\,\partial^{\mu_1}\phi \ldots \partial^{\mu_d} \phi)\,.
\end{equation}
Such a theory correspons to $v=d+1$, $\sigma=1$ and $\rho=(d-2)/(d-1)$.

\subsection*{Dirac-Born-Infeld Theory}

The so-called DBI Lagrangian for the single scalar field in $d$-dimensions reads 
\begin{equation}
{\cal L} = -F^d \sqrt{1-\frac{\partial \phi\cdot \partial \phi}{F^d}} +F^d
\end{equation}
The action can be obtained by description of a $d$-brane fluctuating in the $(d+1)$-dimensional spacetime with a flat metric $\text{diag}(\eta_{\alpha\beta},-1)$. As a consequence this theory must be invariant under the shift symmetry and  $(d+1)$-dimensional Lorentz symmetry
\begin{equation}
\phi \to \phi + a + \theta\cdot x - F^{-d} \theta \cdot \phi(x) \partial \phi(x)\,.
\end{equation}
DBI corresponds to the theory with $\sigma=2$ and $\rho=1$.

\subsection*{$P(X)$ Theory}
The DBI discussed above can be considered a special case of a general class of theories,
\begin{equation}
{\cal L} =  F^d P\Bigl(\frac{\partial \phi\cdot \partial \phi}{F^d}\Bigr)\,,
\end{equation}
occasionally referred to in the context of inflaton cosmology as $P(X)$ theories.  Here $P$ is a Taylor expansion of the form $P(x)= \tfrac12 x + {\cal O}(x^2)$.  This theory is manifestly invariant under the shift symmetry $\phi \to \phi+a$ and thus exhibits $\sigma=1$ and $\rho=1$.  This soft behavior is trivial, since the soft degree matches the number of derivatives per field.

\subsection*{Galileon}

Lagrangian of the so-called Galileon in $d$-dimension consists of $d+1$ terms
\begin{equation}
\mathcal{L}=\sum_{n=1}^{d+1}d_{n}\phi \mathcal{L}_{n-1}^{\mathrm{der}}\,,
\label{eq:galileon}
\end{equation}
with the total derivative term at valency $n$, explicitly given by
\begin{equation}
\mathcal{L}_{n}^{\mathrm{der}}=\varepsilon ^{\mu _{1}\ldots \mu
_{d}}\varepsilon ^{\nu _{1}\ldots \nu _{d}}\prod\limits_{i=1}^{n}\partial
_{\mu _{i}}\partial _{\nu _{i}}\phi \prod\limits_{j=n+1}^{d}\eta _{\mu
_{j}\nu _{j}}=(-1)^{d-1}(d-n)!\det \left\{ \partial ^{\nu _{i}}\partial
_{\nu _{j}}\phi \right\} _{i,j=1}^{n}.  
\end{equation}
For example in $d=4$ we have
\begin{eqnarray}
\mathcal{L}_{0}^{\mathrm{der}} &=&-4!  \notag \\
\mathcal{L}_{1}^{\mathrm{der}} &=&-6\square \phi  \notag \\
\mathcal{L}_{2}^{\mathrm{der}} &=&-2\left[ \left( \square \phi \right)
^{2}-\partial \partial \phi :\partial \partial \phi \right]  \notag \\
\mathcal{L}_{3}^{\mathrm{der}} &=&-\left[ \left( \square \phi \right)
^{3}+2\partial \partial \phi \cdot \partial \partial \phi :\partial \partial
\phi -3\square \phi \partial \partial \phi : \partial \partial \phi %
\right]  \notag \\
\mathcal{L}_{4}^{\mathrm{der}} &=&-\left[ \left( \square \phi \right)
^{4}-6\left( \square \phi \right) ^{2}\partial \partial \phi :\partial
\partial \phi +8\square \phi \partial \partial \phi \cdot \partial \partial
\phi :\partial \partial \phi \right.  \notag \\
&&\left. -6\partial \partial \phi \cdot \partial \partial \phi \cdot
\partial \partial \phi :\partial \partial \phi +3\left( \partial \partial
\phi :\partial \partial \phi \right) ^{2}\right] .
\end{eqnarray}
This Lagrangian has a lowest interaction term with valency 3, but as shown in \cite{Kampf:2014rka} we can always remove it using a duality transformation, which doesn't change the structure of other vertices.
The Galileon Lagrangian represent the most general theory for single scalar whose equation of motion involves just the second derivatives of the field and is invariant under the Galilean symmetry 
\begin{equation}
\phi \to \phi + a + b\cdot x\,.
\end{equation}
According to the soft theorem this theory has $\sigma=2$ and $\rho=2$.

\subsection*{Special Galileon}

In \cite{Cheung:2014dqa} it was found that the Galileon with the 4pt interaction term in $d=4$ has even stronger soft limit behavior than naively predicted by the symmetry argument. In fact, $A_n \sim {\cal O}(p^3)$ rather than just $A_n \sim {\cal O}(p^2)$. This was a signal for a hidden symmetry which was indeed discovered shortly after in \cite{Hinterbichler:2015pqa}.
The special Galileon can be obtained from (\ref{eq:galileon}) with 
\begin{equation}
d_{2n} = \frac{(-1)^d}{(2n)! (d-2n+1)!}\, \frac{1}{\alpha^{2(n-1)}}\,,\qquad d_{2n+1}=0\,.
\end{equation}
In the case of four dimension there is only one interaction term
\begin{equation}
{\cal L}_{int} = \frac{1}{4!} \frac{1}{\alpha^2}\phi {\cal L}_3^{\rm der}\,.
\end{equation}
The hidden symmetry is given by
\begin{equation}
\phi \to \phi + \theta^{\mu\nu} (\alpha^2 x_\mu x_\nu - \partial_\mu \phi \partial_\nu \phi)\,.
\end{equation}
According our definition this means that $\sigma=3$ and $\rho=2$.

\subsection*{Multi-field Galileon}

There are at least two posibilities how to generalize the Galileon
Lagrangian for scalar multiplet. The first one is a straightforward $U(N)$ symmetric generalization of the $n-$point interaction
term
\[
\mathcal{L}_{n}=\varepsilon ^{\mu _{1}\ldots \mu _{d}}\varepsilon ^{\nu
_{1}\ldots \nu _{d}}\mathrm{Tr}\left( \phi \partial _{\mu _{1}}\partial
_{\nu _{1}}\phi \ldots \partial _{\mu _{n}}\partial _{\nu _{n}}\phi \right)
\prod\limits_{j=n+1}^{d}\eta _{\mu _{j}\nu _{j}}
\]
where $\phi =\phi ^{a}T^{a}$ and $T^{a}$ are the generators of $U(N)$. The corresponding action is invariant with respect to the linear shift symmetry and the $U(N)$ symmetry
\begin{eqnarray*}
\phi ^{a} &\rightarrow &\phi ^{a}+b^{a}+c^{a}\cdot x \\
\phi  &\rightarrow &U\phi U^{+},~~U\in SU\left( N\right) 
\end{eqnarray*}
which is responsible for the ${\cal O}( p^{2}) $ soft behavior of the
scattering amplitudes. Moreover, because of the single trace structure of
the interaction terms, the full amplitudes can be flavor-ordered and cyclically
ordered Feynman rules can be formulated. Of course we could also include
interaction terms with multiple traces without spoiling the symmetry and
soft limit properties, e.g.%
\begin{eqnarray*}
\mathcal{L}_{n,k_{1},\ldots ,k_{m}=d} &=&\varepsilon ^{\mu _{1}\ldots \mu
_{d}}\varepsilon ^{\nu _{1}\ldots \nu _{d}}\prod\limits_{j=n+1}^{d}\eta
_{\mu _{j}\nu _{j}}\mathrm{Tr}\left( \phi \partial _{\mu _{1}}\partial _{\nu
_{1}}\phi \ldots \partial _{\mu _{k_{1}}}\partial _{\nu _{k_{1}}}\phi
\right)  \\
&&\times \prod\limits_{r=2}^{m}\mathrm{Tr}\left( \partial _{\mu
_{k_{r-1}+1}}\partial _{\nu _{k_{r-1}+1}}\phi \ldots \partial _{\mu
_{k_{r}}}\partial _{\nu _{k_{r}}}\phi \right) ,
\end{eqnarray*}%
however then the usual stripping of the amplitudes is not possible.

Another generalization follows the brane construction described in
\cite{Hinterbichler:2015pqa}. Such generalization has naturally a $O(N)$
symmetry as a remnant of the Lorentz symmetry of the $d+N$ dimensional
target space in which the $d-$dmesional brane propagates. As shown in \cite{Hinterbichler:2015pqa}, there are only even-point vertices allowed by symmetry and
the $2n-$point Lagrangian has the general form 
\begin{eqnarray*}
\mathcal{L}_{2n} &=&\varepsilon ^{\mu _{1}\ldots \mu _{d}}\varepsilon ^{\nu
_{1}\ldots \nu _{d}}\prod\limits_{j=2n+1}^{d}\eta _{\mu _{j}\nu _{j}} \\
&&\times \sum\limits_{a_{i}=1}^{N}\phi ^{a_{1}}\partial _{\mu _{1}}\partial
_{\nu _{1}}\phi ^{a_{1}}\partial _{\mu _{2}}\partial _{\nu _{2}}\phi
^{a_{2}}\partial _{\mu _{3}}\partial _{\nu _{3}}\phi ^{a_{2}}\ldots \partial
_{\mu _{2n-1}}\partial _{\nu _{2n-1}}\phi ^{a_{n}}\partial _{\mu
_{2n}}\partial _{\nu _{2n}}\phi ^{\alpha _{n}}
\end{eqnarray*}%
The action is invariant with respect to the linear shift symmetry and the $O(N)$ symmetry
\begin{eqnarray*}
\phi ^{a} &\rightarrow &\phi ^{a}+b^{a}+c^{a}\cdot x \\
\phi ^{a} &=&R_{b}^{a}\phi ^{b},~~R\in O(N)
\end{eqnarray*}
and thus the ${\cal O}( p^{2}) $ soft limit is guaranteed. This
generalization does not allow for the usual stripping of the amplitudes.

\subsection*{Multi-field DBI}

The natural generalization of the single scalar DBI Lagrangian can be
obtained as the lowest order action of the $d-$ dimensional brane
propagating in $d+N$ dimensional flat space. The embeding of the brane is
described by%
\[
X^{A}=Y^{A}\left( \xi \right) 
\]%
where $A=0,1,\ldots ,d+N-1$, and the parameters are $\xi \equiv \xi ^{\mu }$
where $\mu =0,\ldots ,d-1$. The induced metric on the brane is%
\[
\mathrm{d}s^{2}=\eta _{AB}\partial _{\mu }Y^{A}\partial _{\nu }Y^{B}\mathrm{d%
}\xi ^{\mu }\mathrm{d}\xi ^{\nu }\equiv g_{\mu \nu }\mathrm{d}\xi ^{\mu }%
\mathrm{d}\xi ^{\nu } 
\]%
and the leading order reparameterization invariant action reads%
\[
S=-F^{d}\int \mathrm{d}^{d}\xi \sqrt{\left( -1\right) ^{d-1}\det \left(
g_{\mu \nu }\right) }=-F^{d}\int \mathrm{d}^{d}\xi \sqrt{\left( -1\right)
^{d-1}\det \left( \eta _{AB}\partial _{\mu }Y^{A}\partial _{\nu
}Y^{B}\right) } 
\]%
where $\ F$ is a constant with $\dim F=1$. Let us fix new parameterization
in terms of parameters $x^{\mu }~$where%
\[
x^{\mu }=Y^{\mu }\left( \xi \right) 
\]%
and denote%
\[
Y^{d-1+j}\left( \xi \left( x\right) \right) =\frac{\phi ^{j}\left( x\right) 
}{F^{d/2}},~~j=1,\ldots ,N 
\]%
Then%
\[
g_{\mu \nu }=\eta _{\mu \nu }-\frac{1}{F^{d}}\sum_{j}\partial _{\mu }\phi
^{j}\partial _{\nu }\phi ^{j}=\eta _{\mu \alpha }\left( \delta _{\nu
}^{\alpha }-\frac{1}{F^{d}}\eta ^{\alpha \beta }\sum_{j}\partial _{\beta
}\phi ^{j}\partial _{\nu }\phi ^{j}\right) 
\]%
and after some algebra we get%
\[
\sqrt{\left( -1\right) ^{d-1}\det g}=1+\sum_{N=1}^{\infty }\sum_{n=1}^{N}%
\frac{\left( -1\right) ^{n}}{2^{n}n!}\sum_{\sum_{j=1}^{n}m_{j}=N,~m_{j}\geq
1}^{\infty }~\prod\limits_{j=1}^{n}\frac{1}{m_{j}}\mathrm{Tr}\alpha ^{m_{j}} 
\]%
where the $N\times N$ matrix $\alpha $ is defined as%
\[
\alpha ^{ij}=\frac{1}{F^{d}}\partial \phi ^{i}\cdot \partial \phi ^{j}. 
\]%
We get then for the first three terms of the Lagrangian%
\begin{eqnarray*}
\mathcal{L}_{N=1} &=&~~\frac{1}{2}\mathrm{Tr}\alpha =\sum_{i}\frac{1}{2}%
\partial \phi ^{i}\cdot \partial \phi ^{i} \\
\mathcal{L}_{N=2} &=&~~\frac{1}{2^{1}}\frac{\mathrm{Tr}\alpha ^{2}}{2}-~%
\frac{1}{2^{2}2!}\left( \mathrm{Tr}\alpha \right) ^{2}=\frac{1}{4F^{d}}%
\sum_{j,i}\left( \partial \phi ^{i}\cdot \partial \phi ^{j}\partial \phi
^{j}\cdot \partial \phi ^{i}-\frac{1}{2}\partial \phi ^{i}\cdot \partial
\phi ^{i}\partial \phi ^{j}\cdot \partial \phi ^{j}\right) \\
\mathcal{L}_{N=3} &=&\frac{1}{2}\frac{\mathrm{Tr}\alpha ^{3}}{3}-\frac{1}{4}%
\mathrm{Tr}\alpha \frac{\mathrm{Tr}\alpha ^{2}}{2}+\frac{1}{48}\left( 
\mathrm{Tr}\alpha \right) ^{3} \\
&=&\frac{1}{F^{2d}}\sum_{j,i,k}\left( \frac{1}{6}\partial \phi ^{i}\cdot
\partial \phi ^{j}\partial \phi ^{j}\cdot \partial \phi ^{k}\partial \phi
^{k}\cdot \partial \phi ^{i}-\frac{1}{8}\partial \phi ^{k}\cdot \partial
\phi ^{k}\partial \phi ^{i}\cdot \partial \phi ^{j}\partial \phi ^{j}\cdot
\partial \phi ^{i}\right. \\
&&\left. +\frac{1}{48}\partial \phi ^{k}\cdot \partial \phi ^{k}\partial
\phi ^{i}\cdot \partial \phi ^{i}\partial \phi ^{j}\cdot \partial \phi
^{j}\right)
\end{eqnarray*}%
The action is invariant with respect to the linearly realized $O(N)$ flavour
rotations ($\phi ^{i}$ beiing in the defining representation) 
\[
\delta ^{\left( ij\right) }\phi ^{k}=\delta ^{jk}\phi ^{i}-\delta ^{ik}\phi
^{j} 
\]%
and non-linearly realized Minkowski rotations and boosts in the $d+N$
dimensional space 
\begin{eqnarray*}
\delta ^{\left( \alpha j\right) }x^{\mu } &=&\eta ^{\mu \alpha }\frac{\phi
^{j}}{F^{d/2}} \\
\delta ^{\left( \alpha j\right) }\phi ^{k} &=&F^{d/2}x^{\alpha }\delta ^{jk}.
\end{eqnarray*}%
The latter symmetry is responsible for the ${\cal O}( p^{2}) $ soft
limit of the scattering amplitudes. However, the structure of the Lagrangian
does not allow for introduction of flavor-ordered amplitudes.

\bibliographystyle{JHEP}
\bibliography{simpEFT}

\end{document}